\documentclass[journal,onecolumn]{IEEEtran}
\pdfoutput=1

\usepackage[utf8]{inputenc}
\usepackage[T1]{fontenc}
\usepackage[english]{babel}
\usepackage{graphicx,grffile}

\newif\ifarxiv\arxivtrue
\newif\ifarxivstrict\arxivstrictfalse
\ifarxiv
  \arxivstricttrue 
\fi
\newif\ifsubmit\submitfalse
\ifsubmit\else
  \usepackage{hyperref} 
  \ifarxiv\else
    \usepackage{orcidlink} 
  \fi
  \newcommand{\tumail}[1]{\href{mailto:#1@tum.de}{#1}}
  \hypersetup{
   unicode,
   psdextra,
   pdfauthor={Johannes Rosenberger, Uzi Pereg, and Christian Deppe},
   pdftitle={Identification over Compound Multiple-Input Multiple-Output Broadcast Channels},
   pdfkeywords={
     Identification capacity, compound channels, broadcast communication, multiple-input multiple-output
   },
   pdfsubject={},
   pdflang={English}}
\fi

\providecommand{\tumail}[1]{#1}
\providecommand{\href}[2]{#2}
\providecommand{\texorpdfstring}[2]{#1}
\providecommand{\orcidlink}[1]{}

\usepackage{amsmath,mathtools}
\usepackage{amssymb}
\usepackage{amsthm}
\usepackage{thmtools}

\usepackage{mathrsfs}

\usepackage{calc}

\usepackage[square,numbers,sort&compress]{natbib}
\ifarxivstrict\else
\fi

\newif\iftumcd\tumcdfalse 
\usepackage{xinttools}
\usepackage{tikz}
\usetikzlibrary{math}
\def\zChannelStateA{0.1}
\def\zChannelStateB{0.5}
\usetikzlibrary{fpu}
\pgfkeys{/pgf/fpu}
\pgfkeys{/pgf/fpu/output format=fixed}
\tikzmath{
  function fold(\p, \q) { return \p*(1-\q) + \q*(1-\p); };
  function Hbin(\p) { return - \p * log2(\p) - (1 - \p)*log2(1-\p); };
  function Ibin(\p, \q) { return Hbin(\p) - Hbin(\q)); };
  function Igauss(\s) { return log2(1 + \s)/2; };
  function IzChannel(\p, \d) { return Hbin(\p * \d) - \p * Hbin(\d); };
  function zChannelPopt(\d) {
    real \g;
    let \g = (1-\d)^((1-\d)/\d);
    return \g/(1 + \d*\g);
  };
}
\pgfkeys{/pgf/fpu=false}

\IfFileExists{tex/shorthands.tex}{\usepackage{bm}
\usepackage{xparse}

\def\cA{{\mathcal{A}}}
\def\cB{{\mathcal{B}}}
\def\cC{{\mathcal{C}}}
\def\cD{{\mathcal{D}}}

\def\cG{{\mathcal{G}}}

\def\cL{{\mathcal{L}}}

\def\cN{{\mathcal{N}}}

\def\cP{{\mathcal{P}}}

\def\cS{{\mathcal{S}}}
\def\cT{{\mathcal{T}}}

\def\cV{{\mathcal{V}}}
\def\cW{{\mathcal{W}}}
\def\cX{{\mathcal{X}}}
\def\cY{{\mathcal{Y}}}

\def\scrR{{\mathscr{R}}}
\def\scrS{{\mathscr{S}}}

\bmdefine\ba{a}
\bmdefine\bb{b}
\bmdefine\bc{c}
\bmdefine\bd{d}
\bmdefine\be{e}
\bmdefine\bff{f}
\bmdefine\bg{g}
\bmdefine\bh{h}
\bmdefine\bi{i}
\bmdefine\bj{j}
\bmdefine\bk{k}
\bmdefine\bl{l}
\bmdefine\bmm{m}
\bmdefine\bn{n}
\bmdefine\bo{o}
\bmdefine\bp{p}
\bmdefine\bq{q}
\bmdefine\br{r}
\bmdefine\bs{s}
\bmdefine\bt{t}
\bmdefine\bu{u}
\bmdefine\bv{v}
\bmdefine\bw{w}
\bmdefine\bx{x}
\bmdefine\by{y}
\bmdefine\bz{z}

\bmdefine\bA{A}
\bmdefine\bB{B}
\bmdefine\bC{C}
\bmdefine\bD{D}
\bmdefine\bE{E}
\bmdefine\bF{F}
\bmdefine\bG{G}
\bmdefine\bH{H}
\bmdefine\bI{I}
\bmdefine\bJ{J}
\bmdefine\bK{K}
\bmdefine\bL{L}
\bmdefine\bM{M}
\bmdefine\bN{N}
\bmdefine\bO{O}
\bmdefine\bP{P}
\bmdefine\bQ{Q}
\bmdefine\bR{R}
\bmdefine\bS{S}
\bmdefine\bT{T}
\bmdefine\bU{U}
\bmdefine\bV{V}
\bmdefine\bW{W}
\bmdefine\bX{X}
\bmdefine\bY{Y}
\bmdefine\bZ{Z}

\bmdefine\bsA{A}
\bmdefine\bsB{B}
\bmdefine\bsC{C}
\bmdefine\bsD{D}
\bmdefine\bsE{E}
\bmdefine\bsF{F}
\bmdefine\bsG{G}
\bmdefine\bsH{H}
\bmdefine\bsI{I}
\bmdefine\bsJ{J}
\bmdefine\bsK{K}
\bmdefine\bsL{L}
\bmdefine\bsM{M}
\bmdefine\bsN{N}
\bmdefine\bsO{O}
\bmdefine\bsP{P}
\bmdefine\bsQ{Q}
\bmdefine\bsR{R}
\bmdefine\bsS{S}
\bmdefine\bsT{T}
\bmdefine\bsU{U}
\bmdefine\bsV{V}
\bmdefine\bsW{W}
\bmdefine\bsX{X}
\bmdefine\bsY{Y}
\bmdefine\bsZ{Z}

\def\bcD{\bm{\mathcal{D}}}

\def\bcL{\bm{\mathcal{L}}}

\def\bcV{\bm{\mathcal{V}}}

\def\tA{\tilde{A}}
\def\tB{\tilde{B}}
\def\tC{\tilde{C}}
\def\tD{\tilde{D}}
\def\tE{\tilde{E}}
\def\tF{\tilde{F}}
\def\tG{\tilde{G}}
\def\tH{\tilde{H}}
\def\tI{\tilde{I}}
\def\tJ{\tilde{J}}
\def\tK{\tilde{K}}
\def\tL{\tilde{L}}
\def\tM{\tilde{M}}
\def\tN{\tilde{N}}
\def\tO{\tilde{O}}
\def\tP{\tilde{P}}
\def\tQ{\tilde{Q}}
\def\tR{\tilde{R}}
\def\tS{\tilde{S}}
\def\tT{\tilde{T}}
\def\tU{\tilde{U}}
\def\tV{\tilde{V}}
\def\tW{\tilde{W}}
\def\tX{\tilde{X}}
\def\tY{\tilde{Y}}
\def\tZ{\tilde{Z}}

\bmdefine\btA{\tilde{A}}
\bmdefine\btB{\tilde{B}}
\bmdefine\btC{\tilde{C}}
\bmdefine\btD{\tilde{D}}
\bmdefine\btE{\tilde{E}}
\bmdefine\btF{\tilde{F}}
\bmdefine\btG{\tilde{G}}
\bmdefine\btH{\tilde{H}}
\bmdefine\btI{\tilde{I}}
\bmdefine\btJ{\tilde{J}}
\bmdefine\btK{\tilde{K}}
\bmdefine\btL{\tilde{L}}
\bmdefine\btM{\tilde{M}}
\bmdefine\btN{\tilde{N}}
\bmdefine\btO{\tilde{O}}
\bmdefine\btP{\tilde{P}}
\bmdefine\btQ{\tilde{Q}}
\bmdefine\btR{\tilde{R}}
\bmdefine\btS{\tilde{S}}
\bmdefine\btT{\tilde{T}}
\bmdefine\btU{\tilde{U}}
\bmdefine\btV{\tilde{V}}
\bmdefine\btW{\tilde{W}}
\bmdefine\btX{\tilde{X}}
\bmdefine\btY{\tilde{Y}}
\bmdefine\btZ{\tilde{Z}}

\bmdefine\tbA{\tilde{A}}
\bmdefine\tbB{\tilde{B}}
\bmdefine\tbC{\tilde{C}}
\bmdefine\tbD{\tilde{D}}
\bmdefine\tbE{\tilde{E}}
\bmdefine\tbF{\tilde{F}}
\bmdefine\tbG{\tilde{G}}
\bmdefine\tbH{\tilde{H}}
\bmdefine\tbI{\tilde{I}}
\bmdefine\tbJ{\tilde{J}}
\bmdefine\tbK{\tilde{K}}
\bmdefine\tbL{\tilde{L}}
\bmdefine\tbM{\tilde{M}}
\bmdefine\tbN{\tilde{N}}
\bmdefine\tbO{\tilde{O}}
\bmdefine\tbP{\tilde{P}}
\bmdefine\tbQ{\tilde{Q}}
\bmdefine\tbR{\tilde{R}}
\bmdefine\tbS{\tilde{S}}
\bmdefine\tbT{\tilde{T}}
\bmdefine\tbU{\tilde{U}}
\bmdefine\tbV{\tilde{V}}
\bmdefine\tbW{\tilde{W}}
\bmdefine\tbX{\tilde{X}}
\bmdefine\tbY{\tilde{Y}}
\bmdefine\tbZ{\tilde{Z}}

\def\tcA{\mathcal{\tA}}
\def\tcB{\mathcal{\tB}}
\def\tcC{\mathcal{\tC}}
\def\tcD{\mathcal{\tD}}
\def\tcE{\mathcal{\tE}}
\def\tcF{\mathcal{\tF}}
\def\tcG{\mathcal{\tG}}
\def\tcH{\mathcal{\tH}}
\def\tcI{\mathcal{\tI}}
\def\tcJ{\mathcal{\tJ}}
\def\tcK{\mathcal{\tK}}
\def\tcL{\mathcal{\tL}}
\def\tcM{\mathcal{\tM}}
\def\tcN{\mathcal{\tN}}
\def\tcO{\mathcal{\tO}}
\def\tcP{\mathcal{\tP}}
\def\tcQ{\mathcal{\tQ}}
\def\tcR{\mathcal{\tR}}
\def\tcS{\mathcal{\tS}}
\def\tcT{\mathcal{\tT}}
\def\tcU{\mathcal{\tU}}
\def\tcV{\mathcal{\tV}}
\def\tcW{\mathcal{\tW}}
\def\tcX{\mathcal{\tX}}
\def\tcY{\mathcal{\tY}}
\def\tcZ{\mathcal{\tZ}}

\bmdefine\tbcA{\tcA}
\bmdefine\tbcB{\tcB}
\bmdefine\tbcC{\tcC}
\bmdefine\tbcD{\tcD}
\bmdefine\tbcE{\tcE}
\bmdefine\tbcF{\tcF}
\bmdefine\tbcG{\tcG}
\bmdefine\tbcH{\tcH}
\bmdefine\tbcI{\tcI}
\bmdefine\tbcJ{\tcJ}
\bmdefine\tbcK{\tcK}
\bmdefine\tbcL{\tcL}
\bmdefine\tbcM{\tcM}
\bmdefine\tbcN{\tcN}
\bmdefine\tbcO{\tcO}
\bmdefine\tbcP{\tcP}
\bmdefine\tbcQ{\tcQ}
\bmdefine\tbcR{\tcR}
\bmdefine\tbcS{\tcS}
\bmdefine\tbcT{\tcT}
\bmdefine\tbcU{\tcU}
\bmdefine\tbcV{\tcV}
\bmdefine\tbcW{\tcW}
\bmdefine\tbcX{\tcX}
\bmdefine\tbcY{\tcY}
\bmdefine\tbcZ{\tcZ}

\def\RR{\mathbb{R}}

\ifx\setmathfont\undefined
  \DeclareSymbolFont{bbold}{U}{bbold}{m}{n}
  \DeclareSymbolFontAlphabet{\mathbbm}{bbold}
  \def\ind{\mathbbm{1}\tup} 
\else
  \def\ind{\mathbb{1}\tup} 
\fi

\let\expect\EE

\DeclareMathOperator*{\argmax}{arg\:max}
\DeclareMathOperator*{\argmin}{arg\:min}

\let\transp\intercal
\DeclareMathOperator{\trace}{trace}
\DeclareMathOperator{\diag}{diag}

\def\DeclareGroup#1#2#3{%
  \DeclareDocumentCommand{#1}{om}{%
     \IfValueTF{##1}{##1#2##2##1#3}{\left#2##2\right#3}%
  }%
}
\DeclareGroup{\ceil}{\lceil}{\rceil}
\DeclareGroup{\floor}{\lfloor}{\rfloor}
\DeclareGroup{\intv}{[}{]}
\DeclareGroup{\abs}||
\DeclareGroup{\set}\{\}
\DeclareGroup{\tup}()
\DeclareGroup{\sprod}{\langle}{\rangle}

\usepackage{amsthm,thmtools}

\DeclareRobustCommand{\declarefallbackthm}[2][]{
  \ifcsname #2\endcsname\else\declaretheorem[#1]{#2}\fi
}

\declarefallbackthm[name=Proposition]{prop}
\declarefallbackthm{lemma}
\declarefallbackthm{corollary}
\declarefallbackthm{theorem}
\newcounter{sublemma}[lemma]

\def\sublemma{\refstepcounter{sublemma}\item[\alph{sublemma})]}
\declarefallbackthm{claim}
\declarefallbackthm[style=definition]{definition}
\declarefallbackthm[style=remark]{remark}
\declarefallbackthm[name=Example]{exmpl}

\edef\colorT{\iftumcd TUMRed\else red\fi}
\edef\colorID{\iftumcd TUMBlue\else blue\fi}
\edef\colorIDo{\iftumcd TUMRed\else red\fi}

\def\Rpool{R_{\text{pool}}}

\def\Xpool{F}

\def\Rbin{\tilde{R}}

\makeatletter
\DeclareDocumentCommand{\jorsn@enc@template}{O{Q}moo}%
  {\IfValueF{#3}{#2{#1}}{#2{#1}_{#3}}\IfValueT{#4}{\tup{#4}}}
\let\encTpl\jorsn@enc@template
\def\enc{\jorsn@enc@template\empty}
\def\encAlt{\jorsn@enc@template\tilde}

\def\Enc{\jorsn@enc@template\bm}
\def\EncAlt{\jorsn@enc@template[\tilde{Q}]\bm}

\makeatother

\def\err{e}
\def\erra{\err_{1}}
\def\errb{\err_{2}}

\let\saerr\aerr
\def\saerrya{\saerr_{1,1}}
\def\saerryb{\saerr_{1,2}}
\def\saerrza{\saerr_{2,1}}
\def\saerrzb{\saerr_{2,2}}

\def\codet{\mathcal{C}}

\def\codeBL{\psi}
\def\CodeBL{\Psi}

\let\type\hat

\def\capT{\mathsf{C}_\mathrm{T}}

\def\capID{\mathsf{C}_\mathrm{ID}}
\def\capSAID{\mathsf{C}_\mathrm{ID}}

}{}

\author{%
  \IEEEauthorblockN{%
    Johannes~Rosenberger\ifarxiv\else\orcidlink{0000-0003-2267-3794}\fi, %
    Uzi~Pereg\ifarxiv\else\orcidlink{0000-0002-3259-6094}\fi,%
      ~\IEEEmembership{Member,~IEEE} %
    and Christian~Deppe\ifarxiv\else\orcidlink{0000-0002-2265-4887}\fi
      ,~\IEEEmembership{Member,~IEEE} %
  } %
  \thanks{A part of this paper was submitted to the %
    2022 IEEE International Conference on Communications (ICC).} %
  \thanks{%
    The authors are with the
    Institute for Communications Engineering,
    Department of Electrical and Computer Engineering,
    Technical University of Munich, Germany
    (e-mail: \{\tumail{johannes.rosenberger},\tumail{uzi.pereg},\tumail{christian.deppe}\}@tum.de).} %
}
\title{Identification over Compound Multiple-Input \\ Multiple-Output Broadcast Channels}

\begin{document}

\pgfkeys{/pgf/number format=fixed}
\pgfkeys{/pgf/number format/precision=4} 

\maketitle

\begin{abstract}
The identification capacity region of the compound broadcast channel
is determined under an average error criterion, where
the sender has no channel state information.
We give single-letter identification capacity formulas for discrete channels
and multiple-input multiple-output Gaussian channels
under an average input constraint.
The capacity theorems apply to general discrete memoryless broadcast channels.
This is in contrast to the transmission setting,
where the capacity is only known for special cases, notably the degraded broadcast channel and the multiple-input multiple-output
broadcast channel with private messages.
Furthermore, the identification capacity region of the compound multiple-input multiple-output broadcast channel
can be larger than the transmission capacity region.
This is a departure from the single-user behavior of identification, since
the identification capacity of a single-user channel equals the transmission capacity.
\end{abstract}


\section{Introduction}

\IEEEPARstart{I}{n}
wireless networks nowadays, massive amounts of data are
communicated under challenging conditions \cite{zhangEA2020MIMObeyond5G}.
In particular, fading channels are characterized by fluctuating signal level and uncertain gain \cite{biglieriEA2007mimo_book,lapidothNarayan1998uncertainty,biglieriProakisShamai1998fading_channels}.
The compound setting is a worst case scenario that requires
reliable communication over every possible channel
from an uncertainty set \cite{blackwellBreimanThomasian1959capacityClassOfChannels}. 
Thereby, the compound multiple-input multiple-output broadcast channel (MIMO-BC)
 has raised a lot of attention 
\cite{weingartenLiuShamaiSteinbergViswanath2007compound_DBC_MIMO_isit,weingartenLiuShamaiSteinbergViswanath2009compound_DBC_MIMO,chongLiang2013compound_DBC,chongLiang2014compound_DBC_MIMO_extremal,weingartenShamaiKramer2007compound_MIMO_BC,maddah-ali2009compound_BC_MIMO_DoF,maddah-ali2010compound_BC_MISO_DoF_isit,gouJafarWang2011compound_wireless_network_DoF},
as multiple-antenna transmission allows for substantial
performance gains compared to single-antenna transmission
\cite{telatar1999mimo,foschini1996mimo}.

Unlike Shannon's transmission task \cite{shannon1948it0},
where a transmitter sends a message over a noisy channel,
and the receiver needs to recover the message that was sent,
in some modern event-triggered applications, a receiver performs only a
binary hypothesis test to determine whether a particular message of
interest was sent or not. This setting is known as
identification (ID) via channels \cite{ahlswedeDueck1989id1}.
Possible applications for ID include authentication tasks such
as watermarking
\cite{steinbergMerhav2001id_watermarking,steinberg2002id_watermarking,moulin2001watermarking_information_theory,moulinKoetter2006id_watermark,ahlswedeCai2006watermarking},
private interrogation of devices
\cite{bringerChabanneCohenKindarji2009id_privateInterrogation},
as well as event-driven applications encountered in sensor communication
\cite{guenlueKliewerSchaeferSidorenko2021id_codes},
Industry 4.0 \cite{bocheDeppe2018secureId_wiretap}
and vehicle-to-X communication \cite{bocheDeppe2018secureId_wiretap}.
For example, in vehicle-to-X communication, a vehicle may
announce information about its future movements to surrounding
road users. Every road user is interested in one specific movement
that interferes with its plans, and he checks only if this
movement is announced or not.
This is in contrast to the transmission task, where
every road user has to decode every message,
regardless of their interest.

In practice, often neither sender nor receiver know the exact channel statistics.
The compound channel model is applicable when the channel
variation is sufficiently slow, so that the channel is
approximately constant throughout the transmission
block~\cite{lapidothNarayan1998uncertainty,biglieriProakisShamai1998fading_channels,goldsmith2005wireless_book,biglieriEA2007mimo_book}.
The capacity of the compound discrete memoryless channel was determined by Blackwell
et.\ al.~\cite{blackwellBreimanThomasian1959capacityClassOfChannels} and
Wolfowitz~\cite{wolfowitz1959simultaneousChannels}.
Subsequently, the result was extended to the
Gaussian setting~\cite{rootVaraiya1968compound_Gaussian}.
For compound MIMO channels, the capacity region was optimized under
different fading models and input covariance
constraints~\cite{palomarCioffiLagunas2003compound_mimo_power,wieselEldarYoninaShamai2007mimoCompoundCapacity_optim,denicCharalambousDjouadi2009compoundMimo_bounds,compoundMIMO_additive_uncertainty,loykaCharalambous2015matrixSVD_ineqs,al-aliHo2017MIMO_precoding_optim,al-asadiEA2019worstCase_beamforming}.

A basic model for multi-user communication
is the broadcast channel (BC), where Alice sends messages to Bob and Charly.
The \emph{discrete memoryless broadcast channel} (DMBC) was introduced by Cover
\cite{coverBC} in 1972, but in the general case, the transmission
capacity region is so far not known. The best known lower bound
is due to Marton \cite{marton1979DMBC}, and the best known upper
bound was proven by Nair and El Gamal
\cite{nairElgamal2007BC_outer_bound}.
The two bounds coincide in special cases such as the more capable,
less noisy or degraded DMBC~\cite{elgamal1979BC_class}.
The MIMO BC~\cite{weingartenSteinbergShamai2006gaussian_mimo_BC}
is not necessarily degraded, yet the capacity region
with private messages equals Marton's lower bound.

For compound BCs with perfect channel state information (CSI)
at the receiver, the transmission capacity
region was determined by Weingarten
et.\ al.~\cite{weingartenLiuShamaiSteinbergViswanath2007compound_DBC_MIMO_isit,weingartenLiuShamaiSteinbergViswanath2009compound_DBC_MIMO},
under a degradedness assumption. Both discrete channels and MIMO
Gaussian channels were treated.
Chong and Liang~\cite{chongLiang2013compound_DBC,chongLiang2014compound_DBC-K-receiver,chongLiang2014compound_DBC_MIMO_extremal}
extended the result to discrete BCs and MIMO BCs with
perfect CSI at the receiver, under a weaker degradedness assumption.
Without CSI, Benammar, Piantanida, and Shamai~\cite{benammarPiantanidaShamai2020compound_BC}
derived lower and upper bounds on the capacity region of compound discrete
memoryless and multiple-input single-output BCs,
and they determined the capacity region for special cases of hybrid
binary symmetric/binary erasure BCs.

The ID capacity region of the discrete memoryless channel was determined by
Ahlswede and Dueck~\cite{ahlswedeDueck1989id1,ahlswedeDueck1989id2}.
In general, for single-user channels it equals the transmission capacity
for known channels~\cite{han2003infoSpectrum_book,watanabe2021idMinimaxConverse},
and also for compound channels~\cite{bocheDeppe2018secureId_wiretap} and
for certain continuous channels
\cite{burnashev1999id_approximation_continuous_channel,burnashev2000id_approximation_finite_channel,burnashev2000id_continuous_tit,han2003infoSpectrum_book}
such as MIMO Gaussian channels \cite{labidiDeppeBoche2021ID_mimo}.
However, the ID capacity is a second-order rate, i.e.
the ID code size grows doubly exponentially in the block length,
provided that the encoder has access to a source of randomness.
This is possible by letting the encoding and decoding sets overlap,
and thereby accepting a substantial probability that some
receiver makes an error, if too many receivers are listening at the same time.
For every single receiver, the error probability is still small.
General results for ID are surveyed
in~\cite{han2003infoSpectrum_book,ahlswede2021identification_probabilistic_models}.

The DMBC was also studied for ID, using various error criteria.
Under a maximum-error criterion, Verboven and van der
Meulen~\cite{verbovenMeulen1990idbc}
derived an upper bound on the identification capacity region of the
DMBC, and a lower bound for the DMBC with feedback.
Furthermore, it was shown that the ID capacity region of the
deterministic DMBC is the same, whether feedback is available or not.
Bilik and Steinberg~\cite{bilikSteinberg2001id_dbc} presented bounds
on the ID capacity region of the degraded DMBC.
Ahlswede~\cite{ahlswede2008gtid_updated} proved that the ID capacity
region of a general DMBC with private messages is the same as
with degraded message sets.
It was further shown in~\cite{ahlswede2008gtid_updated} that the ID
capacity region of the DMBC is strictly larger than the
transmission counterpart with degraded message sets.

Bracher and Lapidoth~\cite{bracherLapidoth2017idbc_arxiv,bracher2016PhD}
established the ID capacity region of the DMBC for a semi-average error criterion,
for which the ID messages that the \emph{sender sends} to the two receivers are assumed
to be uniformly distributed.
Since every receiver is interested in one particular message,
his error probabilities are maximized over all messages for this receiver,
but averaged over the message for the other receiver.
In this setting, the ID capacity region of a DMBC \(B\),
from \(X\) to \(Y_1,Y_2\), is given
by~\cite{bracherLapidoth2017idbc_arxiv,bracher2016PhD}
\begin{align}
  \capSAID(B) = \bigcup_{P_X} \set{
    \begin{array}{l l}
      (R_1, R_2) : & R_1 \leq I(X; Y_1), \\
                   & R_2 \leq I(X; Y_2)
    \end{array}
  }.
\end{align}
This holds for the general DMBC, without any channel ordering conditions.
The achievability proof for ID over the BC is based on more advanced methods than the standard Shannon-theoretic argument.
In particular, Bracher and Lapidoth \cite{bracherLapidoth2017idbc_arxiv}
presented a pool-selection code construction
with binning, and bounded the error probabilities
by analyzing the size of the bin intersections.
The converse proof is based on the strong converse for single-user
ID over discrete memoryless channels~\cite{hanVerdu1992idNewResults}.

In this work, we determine the ID capacity region
of the compound BC
under a semi-average error criterion, where
neither sender nor receiver have access to CSI.
We give single-letter ID capacity formulas for discrete channels
under an average input constraint defined by an arbitrary positive function,
and for MIMO Gaussian channels, under an average input power constraint.
CSI at the receiver would result in the same rate region,
since the receiver can learn the state with a short
training sequence \cite[Remark 7.1]{elgamalKim2011network_it}.
Like Bracher and Lapidoth's~\cite{bracherLapidoth2017idbc_arxiv,bracher2016PhD}
result for the DMBC, our results hold for the general compound DMBC and
for the compound MIMO BC, without any channel ordering.
As for the DMBC, the capacity regions of the compound DMBC and compound MIMO BC
can be larger than their transmission counterparts.

As examples, we derive explicit expressions for the ID capacity
regions for symmetric channels, the binary erasure channel and
scalar Gaussian channels.
In those examples, each user can achieve their maximal rate simultaneously.
Thereby, the capacity region is rectangular
and strictly larger than the corresponding transmission capacity.
We also consider the compound broadcast Z-channel with a binary state,
and a Gaussian product channel,
where the ID capacity region is not rectangular, since different
input distributions or transmit power allocations, respectively,
are optimal for the two users.

This paper is organized as follows:
In Section \ref{sec:prelim}, we introduce the notation, define
the communication model, and review basic properties of
typicality and laws of large numbers.
We make some comments about previous results on ID over single-user compound channels.
Section \ref{sec:results} contains our main results.
Examples are treated in \ref{sec:examples}.
Section \ref{sec:proof.capSAID-CBC.achiev} provides the
achievability part of the proof for the capacity region of the
compound discrete BC,
and Section \ref{sec:proof.capSAID-CBC.converse} provides
the converse part.
Section \ref{sec:proof.capSAID-CBC.MIMO}
extends the results from the discrete setting to the MIMO Gaussian setting.
In Section \ref{sec:discussion}, we discuss the implications of our
results and further directions of research.
Finally, the results are summarized in Section \ref{sec:summary}.

\section{Preliminaries}
\label{sec:prelim}

\subsection{The Compound Broadcast Channel}
\label{sec:prelim:channel_model}

A \emph{compound broadcast channel} (CBC) is specified by a family
\(\cB = \set{B_s}_{s \in \cS}\) of discrete memoryless broadcast channels,
indexed by a channel state \(s\in\cS\). Without feedback, the
conditional output distribution has a product form, i.e.
\begin{equation}
  B_s^n(y_1^n,y_2^n|x^n) = \prod_{t=1}^n B_s(y_{1,t}, y_{2,t} | x_t),
\end{equation}
for \(s \in \cS\). Note that the channel state \(s\) is chosen
once at the beginning of the transmission block, and remains constant
throughout the block.
Given a compound BC \(\cB\), the marginal channels of
Receiver 1 and Receiver 2 are defined by
\begin{subequations}
\begin{align}
    W_{1,s}(y_1|x)&= \sum_{y_2 \in \cY_2} B_s(y_1,y_2|x) ,\\
    W_{2,s}(y_2|x)&= \sum_{y_1 \in \cY_1} B_s(y_1,y_2|x) ,
\end{align}
\end{subequations}
respectively, for \(s \in \cS\).
The BC is an extension of the single-user channel
\(W_s : \cX \to \cP(\cY)\). We denote a compound single-user channel by \(\cW = \set{W_s}_{s \in \cS}\).

\begin{remark}
The compound BC is often defined with one state per
receiver~\cite{weingartenLiuShamaiSteinbergViswanath2007compound_DBC_MIMO_isit,weingartenLiuShamaiSteinbergViswanath2009compound_DBC_MIMO,chongLiang2013compound_DBC,chongLiang2014compound_DBC-K-receiver,chongLiang2014compound_DBC_MIMO_extremal},
i.e. $\cB = \set{ (W_{1,s_1}, W_{2,s_2}) : s_1 \in \cS_1,\, s_2 \in \cS_2 }$.
This is a special case of our definition, where \(\cS = \cS_1 \times \cS_2\).
\end{remark}

A \emph{compound MIMO broadcast channel} is described by a
family \(\cB = \set{B(G_{1,s}, G_{2,s})}_{s \in \cS}\) of Gaussian probability density functions (PDFs),
where \(\cS\) is a finite state set and \(G_{k,s} \in \RR^{\rho_{k,s} \times \tau}\),
for the number of transmit antennas \(\tau\) and the number of
receive antennas \(\rho_{k,s}\), at receiver \(k \in \set{1,2}\) and state \(s \in \cS\).
The PDFs are defined such that
\begin{equation}
  \bY_{k,s} = G_{k,s} \bX + \bZ_{k,s},
\end{equation}
where $\bX \in \RR^\tau$, $\bY \in \RR^{\rho_{k,s}}$, and the noise vector \(\bZ_{k,s}\)
follows the multivariate Gaussian distribution \(\cN(0, I_{\rho_{k,s}})\),
for \(k \in \set{1,2}\) and \(s \in \cS\).
In the following, we simply write \(I = I_{\rho_{k,s}}\), since the dimension
is determined by the context.

\begin{remark}[see Remark 9.1 in {\cite{elgamalKim2011network_it}}]
In general, a MIMO Gaussian channel can always be transformed to have unit noise covariance:
Consider the noise \(\bZ_{k,s} \sim \cN(0, \Sigma_{k,s})\) and \(\Sigma_{k,s} \neq I\).
Note that $\Sigma_{k,s}$ must be strictly positive definite, since
$f_{\bZ_{k,s}}(\bz) = \abs{\Sigma_{k,s}}^{-\frac{1}{2}} \exp\tup{\bz^\transp \Sigma_{k,s}^{-1} \bz}$.
and therefore $\Sigma_{k,s}$ must be invertible. Hence,
\(\Sigma_{k,s}^{-\frac{1}{2}} \bZ_{k,s} \sim \cN(0, I)\) and the receiver can
postprocess the output to transform the channel into
\begin{equation}
  \tbY_{k,s} = \Sigma_{k,s}^{-\frac{1}{2}} G_{k,s} \bX + \Sigma_{k,s}^{-\frac{1}{2}} \bZ_{k,s}.
\end{equation}
\label{remark:gaussian.MIMO.covariance}
\end{remark}

\subsection{Identification Codes}
\label{sec:prelim:ID}

In the following, we define the communication task of identification over compound channels,
where the decoder is not required to recover the sender's message
\(i\), but simply determines whether a particular message \(i'\) was sent or not.
\begin{definition}
An \(\tup{N_1, N_2, n}\) identification code (ID-code) for the
compound BC \(\cB\) is a family of pairs of an encoding PMF
\(\enc[i_1,i_2] \in \cP(\cX^n)\) and decoding sets \(\cD_{k,i_k} \subset \cY_k^n\), for \(i_k \in [N_k]\) and \(k \in \set{1,2}\).
We denote the identification code by
\begin{equation*}
  \codet = \set{ (\enc[i_1,i_2], \cD_{1,i_1}, \cD_{2,i_2}) : i_1 \in [N_1], i_2 \in [N_2] }.
\end{equation*}
Suppose that Receiver~\(k\) is interested in a particular message
\(i'_k \in [N_k]\). He declares that~\(i'_k\) was sent if \(Y_k^n \in \cD_{k,i'_k}\). Otherwise, Receiver~\(k\)
declares that~\(i'_k\) was not sent.

The ID rates of the BC code \(\codet\) are defined as
\(R_k = \frac{1}{n} \log\log(N_k)\) for \(k \in \set{1,2}\).
In this work, we assume that the ID messages \(i_k\) are uniformly distributed over
the set \([N_k]\), for \(k \in \set{1,2}\). Therefore, the error probabilities are defined on
average over the messages for the other receiver.
Furthermore, we consider average input constraints of the form
$\frac{1}{n} \sum_{t=1}^n \gamma(x_t) \leq \Gamma$, where $\gamma : \cX \to [0,\infty)$
can be any non-negative function.
The sender (Alice) makes an error if she transmits
a sequence $X^n$ that doesn't satisfy the constraint.
For $k \in \set{1,2}$, Receiver \(k\) (Bob or Charly) makes an error in one of two cases:
(1) He decides that \(i_k\) was \emph{not} sent (missed ID);
(2) Receiver \(k\) decides that \(i'_k\) was sent,
while in fact \(i_k \neq i'_k\) was sent (false ID). For every \(s \in \cS\),
the probabilites of these three kinds of error,
averaged over all \(i_\ell \in [N_\ell]\), \(\ell \neq k\), are defined as
\begin{subequations}
\label{eq:def.SAID-errProbs}
\begin{align}
  \saerr_{k,0} (n, \codet, i_k)
    &= \sum_{x^n \in \cX^n} \frac{1}{N_\ell} \sum_{i_\ell=1}^{N_\ell}
     \enc[i_1,i_2](x^n) \, \ind[\Big]{\sum_{t=1}^n \gamma(x_t) > n \Gamma }
     \\
  \saerr_{k,1} (B_s, n, \codet, i_k)
    &= \sum_{x^n \in \cX^n} \frac{1}{N_\ell} \sum_{i_\ell=1}^{N_\ell}
     \enc[i_1,i_2](x^n) W^n_{k,s}(\cD_{k,i_k}^c|x^n),
     \\
  \saerr_{k,2} (B_s, n, \codet, i'_k, i_k)
    &= \sum_{x^n \in \cX^n} \frac{1}{N_\ell} \sum_{i_\ell=1}^{N_\ell}
     \enc[i_1,i_2](x^n) W^n_{k,s}(\cD_{k,i'_k}|x^n),
\end{align}
\end{subequations}
For MIMO Gaussian channels, \(\cX\) and $\cY_{k,s}$ are continuous, \(\enc[i_1,i_2]\)
and \(W_{k,s}\) are PDFs, and the sums are replaced integrals.

An \((N_1, N_2, n, \lambda)\) ID-code \(\codet\) for a compound
BC \(\cB\) and input constraint
$\frac{1}{n} \sum_{t=1}^n \gamma(x_t) \leq \Gamma$
satisfies
\begin{subequations}
\begin{align}
   \max_{i_k \in [N_k]} \saerr_{k,0}(n, \codet, i_k) &< \lambda, \\
   \max_{s \in \cS} \max_{i_k \in [N_k]} \saerr_{k,1}(B_s, n, \codet, i_k) &< \lambda, \\
   \max_{s \in \cS} \max_{i_k \in [N_k]} \max_{\substack{i'_k \in [N_k] \\ i'_k \neq i_k}}
     \saerr_{k,2}(B_s, n, \codet, i'_k, i_k) &< \lambda,
\end{align}
\end{subequations}
for \(k \in \set{1,2}\).
An ID rate pair \((R_1, R_2)\) is \emph{achievable} if for every \(\lambda > 0\) and
sufficiently large \(n\), there exists
\linebreak
an \(\tup{\exp{e^{nR_1}}, \exp{e^{nR_2}}, n, \lambda}\) ID-code.
The ID capacity region \(\capSAID(\cB, \Gamma)\) of the compound BC \(\cB\)
under an input constraint $\frac{1}{n} \sum_{t=1}^n \gamma(x_t) \leq \Gamma$
is defined as the set of all achievable ID rate pairs.
The ID capacity without an input constraint is
denoted by $\capSAID(\cB) = \capSAID(\cB, \infty)$.
\label{def:ID-code}
\end{definition}

\begin{remark}
In the definition above,
\(\enc[i_1,i_2]\) may be a \(0\)-\(1\)-distribution, such that we have
deterministic encoding. Usually, in identification, one needs
randomized encoding to achieve code sizes that grow doubly exponentially
in the block length \cite{ahlswedeDueck1989id1,bocheDeppe2018secureId_wiretap}.
However, under a semi-average error criterion, we can view
\(\saerrya, \saerryb\) as error probabilities of a single-user ID-code
\(\set[\big]{(\encAlt[1,i_1], \cD_{1,i_1})}_{i_1=1}^{N_1}\),
with encoding distribution \(\encAlt[1,i_1] = \frac{1}{N_2} \sum_{i_2=1}^{N_2} \enc[i_1,i_2]\),
for \(k \in \set{1,2}\). In the proof of our results, we will use the pool-selection method
of \cite{bracherLapidoth2017idbc_arxiv,bracher2016PhD} to construct $\encAlt[i_1,i_2]$
randomly such that it approximates an encoding distribution for a single-user
ID code for the $k$-th marginal channel.
In this sense, the messages that the sender indends for Receiver 2
are used as randomization for the identification at Receiver 1,
and vice versa \cite{bracherLapidoth2017idbc_arxiv}.
This is only possible if \(\min\set{R_1, R_2} > 0\). Otherwise, we
need stochastic encoding, since we have
a single user setting, then.
\label{remark:determinsticEncoding}
\end{remark}

\subsection{Laws of Large Numbers}
\label{sec:orgc9a398d}
We use basic concepts of typicality and the method of types,
as defined in \cite[Section 2.4]{elgamalKim2011network_it}
and \cite[Chapter 2]{csiszarKoerner2011IT}.
The definitions and lemmas we use are collected in this section.

The \(n\)-type  \(\type{P}_{x^n}\) of a sequence \(x^n \in \cX^n\)
is defined by \(\type{P}_{x^n}(x^n) = \frac{1}{n} \sum_{t=1}^n \ind{x_t=x}\).
The set of all \(n\)-types over a set \(\cX\) is denoted by
\(\cP(n,\cX)\).
Joint and conditional types are defined in a similar manner, as in \cite{elgamalKim2011network_it}.
Furthermore, an \(\epsilon\)-typical set is defined as follows.
Given a PMF  \(P_X \in \cP(\cX)\)  over \(\cX\), define
the \(\epsilon\)-typical set,
\begin{equation*}
\cT_\epsilon^n(P_X) = \set[\big]{x^n \in \cX^n : \abs[\big]{\type{P}_{x^n}(a) - P_X(a)} \leq \epsilon P_X(a),~a \in \cX}.
\end{equation*}
The set of jointly typical sequences
\(\cT_\epsilon^n(P_{XY})\) is defined likewise, where \(\cX\) is replaced by
\(\cX\times{}\cY\).
Given a sequence \(x^n \in \cX^n\), the \emph{conditionally} \(\epsilon\)-typical set with
respect to \(P_{XY}\) is \(\cT_\epsilon^n(P_{XY} | x^n) = \set{y^n \in \cY^n : (x^n, y^n) \in \cT_\epsilon^n(P_{XY})}\).
Basic law-of-large-numbers type properties are given in the lemmas below.
Those are also known as the asymptotic equipartition properties. 

\begin{lemma}
For every \(P_{XY} \in \cP(\cX\times\cY)\), \(\epsilon > 0\) and \(x^n \in \cX^n\)
exists a \(\delta > 0\) such that
\begin{sublemmas}
\sublemma \label{lemma:X_is_untypical_ub}
  $\Pr \tup[\big]{X^n \notin \cT_\epsilon^n(P_X)}
    < 2\abs{\cX}e^{- 2n\delta^2}$
    \cite[Theorem 1.1]{kramer2008multi_user_book},
\sublemma \label{lemma:joint_typicality}
        $\Pr\tup[\big]{ (x^n,Y^n) \in \cT_\epsilon^n(P_{XY}) } \leq 2^{-n[I(X; Y)-2\epsilon H(Y)]}$
    \cite[Theorem 1.3]{kramer2008multi_user_book},
\end{sublemmas}
\end{lemma}

\begin{theorem}[Hoeffding's inequality {\cite[Theorem 1]{hoeffding1963inequalities}}]
Let \(X_t,~t \in [n]\) be a sequence of i.i.d. random variables \(\sim P_X\),
satisfying \(0 < X_t < 1\).
Then for all \(\alpha > 0\),
\begin{equation*}
\Pr\set[\bigg]{\frac{1}{n} \sum_{t=1}^n X_t - \expect [X] \geq \alpha}
  \leq e^{-2 \alpha^2 n}.
  \label{eq:hoeffding}
\end{equation*}
\label{thm:hoeffding}
\end{theorem}

\subsection{Previous Results}
\label{sec:relatedWork}

In the single-user setting, the ID capacity of the single-user
compound channel was determined by Boche and Deppe
\cite{bocheDeppe2018secureId_wiretap}.

\begin{theorem}[see {\cite{bocheDeppe2018secureId_wiretap}}]
The ID capacity of a compound channel \(\cW = \set{W_s}_{s \in \cS}\) is given by
\begin{equation*}
  \capID(\cW) = \max_{P_X \in \cP(\cX)} \min_{s \in \cS} I(X; Y_s),
\end{equation*}
where \(Y_s \sim W_s(\cdot|X)\).
\label{thm:capID.CC}
\end{theorem}

While the explicit proof in \cite{bocheDeppe2018secureId_wiretap} employs
a random binning scheme based on transmission codes \cite{ahlswedeDueck1989id2},
we will provide an an alternative proof by extending the
pool-selection method by Bracher and Lapidoth
\cite{bracherLapidoth2017idbc_arxiv,bracher2016PhD}.
This will enable the same extension to the broadcast setting as in
\cite{bracherLapidoth2017idbc_arxiv,bracher2016PhD}.

We note that the capacity is thus upper-bounded by
the minimum of the capacities of the channels \(W_s\), i.e.
\begin{equation}
  \capID(\cW)
    \leq \min_{s \in \cS} \max_{P_X \in \cP(\cX)} I(X; Y_s)
    = \min_{s \in \cS} \capID(W_s),
    \label{eq:capID_leq_min_capID}
\end{equation}
by the max-min inequality \cite[Eq. 5.46]{boydVandenberghe2004convex}.
Equality holds for certain symmetric examples like the binary symmetric
and the single-user Gaussian channel.
\begin{exmpl}
Represented as a stochastic matrix, the binary symmetric channel is defined by
\begin{equation*}
  \mathrm{BSC}(\delta) =
    \begin{pmatrix}
      1-\delta & \delta \\
      \delta & 1-\delta
    \end{pmatrix}.
\end{equation*}
It is visualized in Figure \ref{fig:bsc}.
\begin{figure*}
\begin{minipage}[b]{.49\textwidth}
  \centering
  \newif\ifqtikz\qtikzfalse


\ifqtikz

\def\capID{\mathsf{C}_{\textrm{ID}}}
\def\capSAID{\mathsf{C}_{\textrm{ID}}}
\def\capT{\mathsf{C}_{\textrm{T}}}

\usetikzlibrary{math}

\input{infoth.tikz}

\edef\colorT{\iftumcd TUMRed\else red\fi}
\edef\colorID{\iftumcd TUMBlue\else blue\fi}
\edef\colorIDo{\iftumcd TUMRed\else red\fi}
\fi

\newcommand{\bscTikz}[1][1]{

\begin{tikzpicture}[auto,inner sep=0.5ex, thick, node distance=2cm and 2cm, box/.style={draw,inner sep=1ex},scale=#1]
  \node (X) at (0,0) {$X$};
  \node (Y) at (4cm,0) {$Y$};
  \matrix[ampersand replacement=\&, row sep=1.5cm, column sep=2.5cm,anchor=west] at (X.east) {
    \node (X1) {$1$}; \& \node (Y1) {$1$}; \\
    \node (X2) {$2$}; \& \node (Y2) {$2$}; \\
  };
  \draw[->] (X1) -- node[above] {$1-\delta$} (Y1);
  \draw[->] (X2) -- node[below] {$1-\delta$} (Y2);
  \draw[->] (X1) -- node[pos=0.33,above] {$\delta$} (Y2);
  \draw[->] (X2) -- node[pos=0.33,below] {$\delta$} (Y1);
\end{tikzpicture}

}

\ifqtikz
\bscTikz
\fi
  \bscTikz
\end{minipage}
\hfill
\begin{minipage}[b]{.49\textwidth}
  \centering
  \newif\ifqtikz\qtikzfalse


\ifqtikz

\def\capID{\mathsf{C}_{\textrm{ID}}}
\def\capSAID{\mathsf{C}_{\textrm{ID}}}
\def\capT{\mathsf{C}_{\textrm{T}}}

\usetikzlibrary{math}

\input{infoth.tikz}

\edef\colorT{\iftumcd TUMRed\else red\fi}
\edef\colorID{\iftumcd TUMBlue\else blue\fi}
\edef\colorIDo{\iftumcd TUMRed\else red\fi}
\fi

\newcommand{\zChannelTikz}[3][]{

\begin{tikzpicture}[auto,inner sep=0.5ex, thick, node distance=2cm and 2cm, Y/.style={#1}]
  \matrix[ampersand replacement=\&, row sep=0.55cm] (X) at (0,0) {
	\node (X=1) {$(1-p)$}; \& \node (X1) {$1$}; \\
    \node {#2}; \\
	\node (X=2) {$(p)$}; \& \node (X2) {$2$}; \\
  };
  \node[Y] (Y) at (5cm,0) {#3};

  \draw[->,Y] (X1) -- node[above] {$1$} +(3.5cm,0) node[right] (Y1) {$1$};
  \draw[->,Y] (X2) -- node[below] {$\epsilon$} +(3.5cm,0) node[right] (Y2) {$2$};
  \draw[->,Y] (X2) -- node[sloped,pos=0.33,above] {$1-\epsilon$} (Y1);

\end{tikzpicture}
}

\ifqtikz
\zChannelTikz[blue,dashed]{X}{Y}
\fi
  \zChannelTikz{$X$}{$Y$}
\end{minipage}

\begin{minipage}[b]{.49\textwidth}
\caption[The Binary Symmetric Channel]{\label{fig:bsc}Binary symmetric channel with crossover probability $\delta$.}
\end{minipage}
\hfill
\begin{minipage}[b]{.49\textwidth}
\caption[The Z-Channel]{\label{fig:z-channel}Z-channel with parameter $\epsilon$.}
\end{minipage}
\end{figure*}

Therefore, the crossover probability is \(\delta\), i.e.
for all \(x,x' \in \set{1,2}\) such that \(x \neq x'\),
\(P_{Y|X}(x|x) = 1-\delta\) and \(P_{Y|X}(x'|x) = \delta\).
Consider \(W_s = \mathrm{BSC}(\delta_s)\) for every \(s \in \cS\). The capacity satisfies
\begin{align}
  \capID(\cW)
    &= \max_{P_X \in \cP(\cX)} \min_{s \in \cS} I(X; Y_s) \nonumber\\
    &= \max_{P_X \in \cP(\cX)} \min_{s \in \cS} \tup{ H(Y_s) - H(Y_s|X) } \nonumber\\
    &\leq 1 - \max_{s \in \cS} H_2(\delta_s),
\end{align}
where \(H_2(\delta) = -\delta \log \delta - (1-\delta) \log (1-\delta)\)
is the binary entropy function.
Note that the inequality is saturated for the input distribution
\(P_X(\cdot) = p = \frac{1}{2}\), for which
\begin{equation}
  P_{Y_s}(y) = p (1-\delta) + (1-p)\delta = \frac{1}{2},
\end{equation}
for all \(s \in \cS\).
Hence, \(H(Y_s) = H_2(\frac{1}{2}) = 1\) and 
\begin{equation}
  \capID(\cW)
    = \min_{s \in \cS} \tup{ 1 - H_2(\delta_s) }
    = \min_{s \in \cS} \capID(W_s).
   \label{eq:capID.single-user.BSC}
\end{equation}
\label{example:BSC.single-user}
\end{exmpl}

This property does not hold in general,
as demonstrated in the next example.

\begin{exmpl}[see Example 7.1 in {\cite{elgamalKim2011network_it}}]
Consider the following Z-channels,
\begin{equation*}
  W_1 =
    \begin{pmatrix}
      1  & 0 \\
      \epsilon & 1-\epsilon
    \end{pmatrix},
    \qquad
  W_2 =
    \begin{pmatrix}
      1-\epsilon & \epsilon \\
      0  & 1
    \end{pmatrix},
\end{equation*}
where \(\epsilon = 1/2\). Let \(\cW = \set{W_1, W_2}\).
The channel \(W_1\) is visualized in Figure \ref{fig:z-channel}.
The cannel \(W_2\) is the same as \(W_1\), but with
reversed order of the input and output alphabet.
The capacity of \(\cW\) is
\begin{align}
  \capID(\cW)
    &= \max_{P_X \in \cP(\cX)} \min_{s \in \cS} I(X; Y_s) \nonumber\\
    &= \max_{p \in [0,1]} H_2(p \epsilon) - p H_2(\epsilon) \nonumber\\
    &= H_2(1/4) - 1/2 H_2(1/2) \nonumber\\
    &= 0.3113,
\end{align}
and it is achieved by \(P_X(\cdot) = \frac{1}{2}\). Thereby, the ID capacity
is \emph{strictly} lower than the ID capacity of the individual channels,
\(W_1, W_2\), as
\begin{equation}
  \capID(W_1) = \capID(W_2) = H(1/5) - 2/5 = 0.3219.
\end{equation}
This is in contrast to Example \ref{example:BSC.single-user} (see \eqref{eq:capID.single-user.BSC}).
\label{example:z-channelElGamalKim}
\end{exmpl}

\section{Results}
\label{sec:results}

Our results are presented below.

\subsection{MIMO Gaussian Channels}
\label{sec:results.mimo}

For a compound MIMO Gaussian BC \(\cB = \set{B(G_{1,s}, G_{2,s})}\),
consider an average input power constraint
$\frac{1}{n} \sum_{t=1}^n \bX_t^\transp \bX_t \leq P$,
i.e. $\gamma(\bX) = \bX^\transp \bX$.
We denote the rate region
\begin{gather}
  \scrR_P(\cB)
    = 
    \bigcup_{\substack{P_1, \dots, P_\tau: \\ \sum_{j=1}^\tau P_j \leq P}}  \set{
    \begin{array}{l l}
     (R_1, R_2) : & \text{For all $k \in \set{1,2}$: } \\
    		  &\displaystyle R_k \leq \min_{s\in\cS} 
       \sum_{j=1}^\tau \frac{1}{2} \log_2 (1 + \lambda_{k,s}^{(j)} P_j)
    \end{array}
    },
    \label{eq:def.R_P}
\end{gather}
where $\lambda_{k,s}^{(1)}, \dots, \lambda_{k,s}^{(\tau)}$ are the eigenvalues
of the matrix $G_{k,s}^\transp G_{k,s}$.

\begin{theorem}
The ID capacity region of the compound MIMO Gaussian broadcast
channel $\cB$
under an average input power constraint $\frac{1}{n} \sum_{t=1}^n \bX_t^\transp \bX_t \leq P$
is given by
\begin{equation*}
  \capSAID(\cB, P) = \scrR_P(\cB).
\end{equation*}
\label{thm:capSAID-CBC.MIMO}
\end{theorem}
\noindent
The proof follows in Section \ref{sec:proof.capSAID-CBC.MIMO}.
Observe that, for every compound MIMO BC, there exists a compound Gaussian Product BC
with the same ID capacity and
\begin{equation}
        \bY_{k, s} = \diag\tup{ \sqrt{\lambda_{k,s}^{(1)}}, \dots, \sqrt{\lambda_{k,s}^{(\tau)}} } \bX + \bZ_{k,s},
  \label{eq:parallelMIMO}
\end{equation}
for every \(k \in \set{1,2}\), \(s \in \cS\) and
\(\bZ_{k,s} \sim \cN(0, I)\).

\subsection{Discrete Channels}
\label{sec:results.discrete}

Consider the compound discrete BC
\(\cB\), as defined in Section \ref{sec:prelim:channel_model},
and an arbitrary non-negative function $\gamma : \cX \to [0, \infty)$.
For this constraint, we denote the rate region
\begin{equation}
  \scrR_\Gamma(\cB) =
  \bigcup_{\substack{
          P_X \in \cP(\cX): \\ \expect[\gamma(X)] \leq \Gamma
  }} \set{
  \begin{array}{l l}
    (R_1, R_2) :
        & \displaystyle R_1 \leq \min_{s\in\cS} I(X; Y_{1,s}), \\
   		& \displaystyle R_2 \leq \min_{s\in\cS} I(X; Y_{2,s})
   	     \end{array}
  }.
  \label{eq:def.R_Gamma}
\end{equation}

\begin{theorem}
The ID capacity region of a CBC \(\cB = \set{B_s : \cX \to \cP(\cY_1 \times \cY_2)}_{s \in \cS}\)
with constraint $\frac{1}{n} \sum_{t=1}^n \gamma(x_t) \leq \Gamma$
satisfies
\begin{equation}
  \capSAID(\cB, \Gamma) = \scrR_\Gamma(\cB).
  \label{eq:thm.capSAID-CBC.bound}
\end{equation}
\label{thm:capSAID-CBC}
\end{theorem}
\noindent
The proof is separated into two parts: In Section \ref{sec:proof.capSAID-CBC.achiev},
we show that all rate pairs in the interior of the region $\scrR_\Gamma(\cB)$
are achievable. In Section \ref{sec:proof.capSAID-CBC.converse} we
show the converse part, i.e. that no rate pair outside $\scrR_\Gamma(\cB)$
can be achieved.


\begin{remark}
By the state approximation of Blackwell, Breiman and Thomasian
\cite[Lemma 4]{blackwellBreimanThomasian1959capacityClassOfChannels},
the results of Theorem \ref{thm:capSAID-CBC} and of Theorem \ref{thm:capSAID-CBC.MIMO}
are immediately extended to arbitrarily large or even non-countable state alphabets.
\label{remark:approximation}
\end{remark}

\begin{remark}
In the single-user setting, the ID and transmission capacity characterizations
are identical. However, in the broadcast ID setting, we see a departure from this
equivalence \cite{ahlswede2008gtid_updated,bracherLapidoth2017idbc_arxiv,bracher2016PhD}.
The examples in the following section demonstrate this departure in a more explicit manner,
showing that the ID capacity region can be strictly larger than the
transmission capacity region. We will come back to this in the sequel.
\end{remark}

\begin{remark}
In general, one cannot necessarily achieve the full capacity of
each marginal channel, i.e.
\begin{align*}
  \scrR_\Gamma(\cB)
  = \set{
    \begin{array}{l l}
      (R_1, R_2) : & R_1 \leq \capID(\cW_1, \Gamma), \\
		   & R_2 \leq \capID(\cW_2, \Gamma)
    \end{array}
  },
\end{align*}
where $\capID(\cW_k, \Gamma) = \max_{P_X : \expect[\gamma(X)] \leq \Gamma} \min_{s \in \cS} I(X; Y_{k,s})$,
since both marginal channels must share the same input distribution.
Equality holds if the same input distribution \(P_X^\star\) maximizes
both minimal mutual informations, i.e. if
\begin{equation}
  P_X^\star = \argmax_{P_X : \expect[\gamma(X)] \leq \Gamma} \min_{s\in\cS} I(X; Y_{1,s})
  = \argmax_{P_X : \expect[\gamma(X)] \leq \Gamma} \min_{s\in\cS} I(X; Y_{2,s}).
  \label{eq:P_X_max_symmetric}
\end{equation}
The examples in Sections \ref{sec:examples.symmetric} and \ref{sec:examples.gaussian} show
that this applies to symmetric channels and to the scalar Gaussian channel,
hence the capacity region is rectangular.
Sections \ref{sec:examples.z-channel} and \ref{sec:examples.MIMO} show that this doesn't hold in
general for broadcast Z-channels and for MIMO channels.
\label{remark:P_X_max_symmetric}
\end{remark}

\section{Examples}
\label{sec:examples}

As examples, we consider the compound symmetric broadcast channels, the compound binary erasure channel,
the compound binary broadcast Z-channel, the compound scalar Gaussian BC,
and a Gaussian product BC.

\subsection{Symmetric Channels}
\label{sec:examples.symmetric}

A channel is called \emph{weakly symmetric} (see \cite[p. 190]{coverThomas2005elements})
if all rows of the transition matrix \(W\) are permutations of each other
and all columns \(y\) have the same sum \(\sum_x W(y|x)\).
The simplest weakly symmetric channel is the binary symmetric channel
\begin{equation*}
  \mathrm{BSC}(\delta) =
    \begin{pmatrix}
      1-\delta & \delta \\
      \delta & 1-\delta
    \end{pmatrix},
\end{equation*}
where the binary input distribution has the form \(P_X = (p, 1-p)\) for some \(p \in [0,1]\).
By Theorem 7.2.1 in \cite{coverThomas2005elements},
every weakly symmetric channel \(W : \cX \to \cP(\cY)\) has
\begin{equation*}
  \max_{P_X \in \cP(\cX)} I(X; Y) = \log\abs\cY - H(Y|X),
\end{equation*}
and the maximizing \(P_X\) is the equidistribution over \(\cX\) (e.g. \(p = 1/2\) for the BSC).
Since the maximizing \(P_X = P_{X^\star}\) is the same for all states \(s \in \cS\) of a compound channel \(\cW\),
by Theorem \ref{thm:capID.CC} the ID capacity of \(\cW\) is given by
\begin{equation}
  \capID(\cW) = \max_{P_X \in \cP(\cX)} \min_{s \in \cS} I(X; Y_s)
  = \min_{s \in \cS} I(X^\star; Y_s)
  = \log\abs\cY - \max_{s \in \cS} H(Y_s|X).
  \label{eq:capID.symmetric}
\end{equation}

Consider a compound broadcast channel \(\cB\) where the marginal channels
\(W_{k,s}\) are weakly symmetric for every receiver \(k \in \set{1,2}\) and state \(s \in \cS\).
Then, we get from Theorem \ref{thm:capSAID-CBC} the capacity region
\begin{align}
  \capSAID(\cB)
  &= \set{
    \begin{array}{l l}
      (R_1, R_2) : & R_1 \leq \capID(\cW_1), \\
		   & R_2 \leq \capID(\cW_2)
    \end{array}
  }.
\end{align}
This is in accordance with Remark \ref{remark:P_X_max_symmetric},
since the same \(P_X(\cdot) = \frac{1}{\abs\cX}\)
maximizes all mutual informations \(I(X; Y_{k,s})\).
Specifically, for \(\cB\) with marginal binary symmetric
channels \(BSC(\delta_s^{(k)})\), we have
\begin{equation}
  \capSAID(\cB) = \set{
    \begin{array}{l l}
      (R_1, R_2) : & R_1 \leq 1 - \max_{s \in \cS} H_2(\delta_s^{(1)}), \\
		   & R_2 \leq 1 - \max_{s \in \cS} H_2(\delta_s^{(2)})
    \end{array}
  }.
\end{equation}
By \cite{elgamalKim2011network_it},
the transmission capacity region of \(\cB\) is
\begin{equation}
  \capT(\cB) = \bigcup_{\alpha \in [0,0.5]} \set{
    \begin{array}{l l}
      (R_1, R_2) : & R_1 \leq \min_{s \in \cS} \tup{ H_2(\alpha \ast \delta_s^{(1)}) - H_2(\delta_s^{(1)}) }, \\
		   & R_2 \leq \min_{s \in \cS} \tup{ 1 - H_2(\alpha \ast \delta_s^{(2)}) }
    \end{array}
  },
\end{equation}
where \(\alpha \ast \delta = \alpha (1-\delta) + \delta (1-\alpha)\).
The capacity regions \(\capSAID(\cB)\) and \(\capT(\cB)\) are visualized in Figure \ref{fig:bsbc.capRegion}. We see that the \(\capID(\cB)\) is rectangular
and strictly larger than \(\capT(\cB)\).

\begin{figure}
  \centering
  \newif\ifqtikz\qtikzfalse


\ifqtikz

\def\capID{\mathsf{C}_{\textrm{ID}}}
\def\capSAID{\mathsf{C}_{\textrm{ID}}}
\def\capT{\mathsf{C}_{\textrm{T}}}

\usetikzlibrary{fpu}
\usetikzlibrary{math}

\input{infoth.tikz}

\edef\colorT{\iftumcd TUMRed\else red\fi}
\edef\colorID{\iftumcd TUMBlue\else blue\fi}
\edef\colorIDo{\iftumcd TUMRed\else red\fi}
\fi

\newcommand{\rateRegionBSBC}[3][20]{

\pgfkeys{/pgf/fpu}
\pgfkeys{/pgf/fpu/output format=fixed}
\tikzmath{
 real \dda, \ddb;
 \dda = #2;
 \ddb = #3;
 function R2(\x) { return Ibin(0.5, fold(\x,\ddb)); };
 function R1(\x) { return Ibin(fold(\x, \dda), \dda); };
}
\pgfkeys{/pgf/fpu=false}

\begin{tikzpicture}[auto,inner sep=1ex, thick, node distance=2cm and 2cm, box/.style={draw,inner sep=1ex}, scale=#1]
\pgfkeys{/pgf/number format=fixed}
\pgfkeys{/pgf/number format/precision=3}
\draw[->] (0,0) -- (0, {Ibin(0.5,\ddb)+0.03}) node[above] {$R_2$};
\draw[->] (0,0) -- ({Ibin(0.5,\dda)+0.03}, 0) node[right] {$R_1$};

\draw[color=\colorID] (0, {Ibin(0.5,\ddb)}) -- +({Ibin(0.5,\dda)}, 0) node[below left] {$\capSAID$} -- ({Ibin(0.5,\dda)}, 0);
\draw (0, {Ibin(0.5,\ddb)}) -- +(-0.008,0) node [left] {$\pgfmathparse{Ibin(0.5,\ddb)}\pgfmathprintnumber\pgfmathresult$};
\draw ({Ibin(0.5,\dda)}, 0) -- +(0,-0.008) node [below]
 {$\pgfmathparse{Ibin(0.5,\dda)}\pgfmathprintnumber\pgfmathresult$};

\draw[thin,color=gray] ({Ibin(0.5,\dda)},0) -- node[sloped,below] {Time Sharing} (0,{Ibin(0.5,\ddb)});
\pgfkeys{/pgf/fpu}
\pgfkeys{/pgf/fpu/output format=fixed}
\draw[dashed,color=\colorT] plot [domain=0:0.5,samples=150] ({R1(\x)}, {R2(\x)});
\pgfkeys{/pgf/fpu=false}
\draw[color=\colorT] ({R1(0.12)}, {R2(0.12)}) node[above] {$\capT$};
\end{tikzpicture}
}

\ifqtikz
\rateRegionBSBC{0.00001}{0.05}
\fi
  \def\bsbcCrossoverA{0.0005}
  \def\bsbcCrossoverB{0.1}
  \rateRegionBSBC[6]{\bsbcCrossoverA}{\bsbcCrossoverB}
  \caption[ID and transmission capacity regions of the compound binary symmetric BC]{%
    \label{fig:bsbc.capRegion}The ID capacity region $\capID(\cB)$
    and transmission capacity region $\capT(\cB)$ of the
    compound binary symmetric broadcast channel $\cB$ with worst crossover
    probabilities $\max_s \delta_s^{(1)} = \bsbcCrossoverA$
    and $\max_s \delta_s^{(2)} = \bsbcCrossoverB$.}
\end{figure}

\subsection{Channels Composed of Symmetric Channels}
\label{sec:examples.composedSymmetric}

Uniform input distribution is not only optimal
for weakly symmetric channels, but also for channels
composed of those, such as the binary erasure channel
\begin{equation*}
  \mathrm{BEC}(\delta) =
    \bordermatrix{
      ~ & 1 & 2 & e \cr
      1 & 1-\delta & 0 & \delta \cr
      2 & 0 & 1-\delta & \delta
      }
      = 
    (1-\delta)
      \underbrace{\begin{pmatrix}
	1 & 0 & 0 \\
	0 & 1 & 0 \\
      \end{pmatrix}}_{W_1}
    + \delta
      \underbrace{\begin{pmatrix}
	0 & 0 & 1 \\
	0 & 0 & 1
      \end{pmatrix}}_{W_2}.
\end{equation*}
Here, the output alphabet is $\cY = \cX \cup \set{e}$,
where $e$ marks an erasure.
Note that both channels \(W_1\) and \(W_2\) are weakly symmetric.
By \cite[p. 189]{coverThomas2005elements},
we have
$\max_{P_X \in \cP(\cX)} I(X; Y) = I(X^\star; Y) = 1-\delta$,
where the uniform input distribution $P_{X^\star}$
is independent of $\delta$.

Consider a compound broadcast channel \(\cB\) where the marginal channels
\(W_{k,s}\) are binary erasure channels
\begin{equation}
  W_{k,s} = \mathrm{BEC}(\delta_s^{(k)}),
\end{equation}
for every \(k \in \set{1,2}\) and \(s \in \cS\).
By Theorem \ref{thm:capID.CC},
the ID capacity of $\cW_k$ is given by
\begin{equation}
  \capID(\cW_k)
    = \max_{P_X \in \cP(\cX)} \min_{s \in \cS} I(X; Y_{k,s})
    = \min_{s \in \cS} I(X^\star; Y_{k,s})
    = 1- \max_{s \in \cS} \delta_s.
\end{equation}
Then, we get from Theorem \ref{thm:capSAID-CBC} the capacity region
\begin{align}
  \capSAID(\cB)
  &= \set{
    \begin{array}{l l}
      (R_1, R_2) : & R_1 \leq \capID(\cW_1), \\
		   & R_2 \leq \capID(\cW_2)
    \end{array}
  }.
\end{align}

\subsection{Broadcast Z-Channel}
\label{sec:examples.z-channel}

Consider the compound broadcast channel \(\cB\) with
the marginal Z-channels
\begin{equation*}
  W_1(\epsilon) =
    \begin{pmatrix}
      1  & 0 \\
      1 - \epsilon & \epsilon
    \end{pmatrix},
\qquad
   W_2(\epsilon) =
     \begin{pmatrix}
       \epsilon  & 1-\epsilon \\
       0 & 1
     \end{pmatrix},
\end{equation*}
for \(\epsilon \in \set{\epsilon_1, \epsilon_2}\). The cannel \(W_2\) is the same as \(W_1\), but with
reversed order of the input and output alphabet.
The resulting broadcast Z-channel \(B(\epsilon)\) is 
visualized in Figure \ref{fig:zBC}.

\begin{figure}
\begin{minipage}[b]{.49\textwidth}
  \centering
    \newif\ifqtikz\qtikzfalse


\ifqtikz

\def\capID{\mathsf{C}_{\textrm{ID}}}
\def\capSAID{\mathsf{C}_{\textrm{ID}}}
\def\capT{\mathsf{C}_{\textrm{T}}}

\usetikzlibrary{math}

\input{infoth.tikz}
\fi

\edef\colorYb{\iftumcd TUMBlue\else blue\fi}

\newcommand{\zBCTikz}{

\begin{tikzpicture}[auto,inner sep=0.5ex, thick, node distance=2cm and 2cm, box/.style={draw,inner sep=1ex}]
  \matrix[ampersand replacement=\&, row sep=1cm] (X) at (0,0) {
	\node (X=1) {$(1-p)$}; \& \node (X1) {$1$}; \\
    \node {$X$}; \\
	\node (X=2) {$(p)$}; \& \node (X2) {$2$}; \\
  };
  \matrix[ampersand replacement=\&, row sep=0.2cm,anchor=west] at (4cm,0) {
    \node (Y11) {$1$}; \\
    \& \node (Y1) {$Y_1$}; \\
    \node (Y12) {$2$}; \\
	\node {\strut}; \\
    \node[color=\colorYb] (Y21) {$1$}; \\
    \& \node[color=\colorYb] (Y2) {$Y_2$}; \\
	\node [color=\colorYb](Y22) {$2$}; \\
  };
  \draw[->] (X1) -- node[above left] {$1$} (Y11);
  \draw[->] (X2) -- node[pos=0.2,below] {$\epsilon$} (Y12);
  \draw[->] (X2) -- node[sloped,pos=0.15,above] {$1-\epsilon$} (Y11);

  \draw[->,color=\colorYb,dashed] (X1) -- node[pos=0.2,above] {$\epsilon$} (Y21);
  \draw[->,color=\colorYb,dashed] (X2) -- node[below left] {$1$} (Y22);
  \draw[->,color=\colorYb,dashed] (X1) -- node[sloped,pos=0.15,below] {$1-\epsilon$} (Y22);

\end{tikzpicture}
}

\ifqtikz
\zBCTikz
\fi
    \zBCTikz
    \caption[The broadcast Z-channel]{\label{fig:zBC}The broadcast Z-channel channel with
     parameter $\epsilon$.}
\end{minipage}
\hfill
\begin{minipage}[b]{.49\textwidth}
  \centering
    \newif\ifqtikz\qtikzfalse


\ifqtikz

\def\capID{\mathsf{C}_{\textrm{ID}}}
\def\capSAID{\mathsf{C}_{\textrm{ID}}}
\def\capT{\mathsf{C}_{\textrm{T}}}

\usetikzlibrary{math}
\usetikzlibrary{fpu}

\input{infoth.tikz}

\edef\colorT{\iftumcd TUMRed\else red\fi}
\edef\colorID{\iftumcd TUMBlue\else blue\fi}
\edef\colorIDo{\iftumcd TUMRed\else red\fi}
\fi

\pgfkeys{/pgf/fpu}
\pgfkeys{/pgf/fpu/output format=fixed}
\tikzmath{
 function R(\p,\d) { return IzChannel(\p,\d); };
 function zp(\d) { return zChannelPopt(\d); };
}
\pgfkeys{/pgf/fpu=false}

\newcommand{\rateRegionZBC}[2][50]{

\pgfkeys{/pgf/fpu}
\pgfkeys{/pgf/fpu/output format=fixed}
\tikzmath{
 real \dda, \ddb, \zp, \Rzp, \Rzpc, \Rb;
 \dda = #2;
 \zp = zp(\dda);
 \Rzp = R(\zp, \dda);
 \Rzpc = R(1-\zp, \dda);
 %
}
\pgfkeys{/pgf/fpu=false}

\begin{tikzpicture}[auto,inner sep=1ex, thick, node distance=2cm and 2cm, scale=#1]
\pgfkeys{/pgf/number format=fixed}
\pgfkeys{/pgf/number format/precision=4}
\draw[->] (0,0) -- (0, {\Rzp + 0.005}) node[above] {$R_2$};
\draw[->] (0,0) -- ({\Rzp+0.005}, 0) node[right] {$R_1$};

\draw[] (0, \Rzp) -- +(-0.002,0) node [left] {\pgfmathprintnumber{\Rzp}};
\draw[](\Rzp, 0) -- +(0,-0.002) node [below] {\pgfmathprintnumber{\Rzp}}; 

\pgfkeys{/pgf/fpu}
\pgfkeys{/pgf/fpu/output format=fixed}
\draw[color=\colorID]
  (0,\Rzp) -- +(\Rzpc,0)
  plot [parametric,domain=\zp:1-\zp,samples=150] ({R(\x,\dda)}, {R(1-\x, \dda)})
  (\Rzp,0) -- +(0, \Rzpc);
\pgfkeys{/pgf/fpu=false}
\draw[color=\colorID] (\Rzp, \Rzpc) node[left] {$\capSAID$};
\end{tikzpicture}
}

\ifqtikz
\rateRegionZBC{0.1}
\fi
    \rateRegionZBC[60]{\zChannelStateA}
    \caption[The capacity region of the compound broadcast Z-channel]{%
      \label{fig:zBC.capRegion}The capacity region of the compound broadcast
       Z-channel with marginal channels $W_k(\epsilon_s)$, for $k,s \in \set{1,2}$ and
       parameters $\epsilon_1 = \zChannelStateA$ and $\epsilon_2 = \zChannelStateB$.}
\end{minipage}
\end{figure}

We derive now the capacity region for the compound BC \(\cB\).
Let the input distribution be \(P_X = (1-p, p)\). Then,
\begin{equation}
  P_{Y_{k,s}} = (1-p\epsilon_s, p\epsilon_s).
\end{equation}
The mutual information for \(W_1(\epsilon_s)\) can be calculated as
\begin{align}
  I(X; Y_{1,s})
    &= H(Y_{1,s}) - H(Y_{1,s}|X) \nonumber\\
    &= H(Y_{1,s}) - \tup[\big]{ (1-p) \cdot 0 + p H_2(1-\epsilon_s) } \nonumber\\
    &= H_2(p\epsilon_s) - p H_2(\epsilon_s).
\end{align}
By \cite{golomb1980z-channel}, the \(p\) that maximizes the mutual information
for \(W_1(\epsilon)\) is given by
\begin{equation}
  p_\epsilon = \frac{\gamma}{1 + \epsilon \gamma},
\end{equation}
where \(\gamma = (1-\epsilon)^{(1-\epsilon)/\epsilon}\).
By symmetry, the mutual information for \(W_2(\epsilon_s)\) is given by
\begin{equation}
  I(X; Y_{2,s}) = H_2((1-p) \epsilon_s) - (1-p) H_2(\epsilon_s).
\end{equation}
This mutual information is maximized for \(p = 1-p_\epsilon\),
and it decreases for decreasing \(\epsilon\). To see this,
consider the derivative
\begin{align}
  \frac{\partial}{\partial \epsilon} \tup{ H_2(p \epsilon) - p H_2(\epsilon) }
    &= \frac{\partial}{\partial \epsilon} \tup[\Big]{
	-p\epsilon \log_2 (p \epsilon) - (1-p\epsilon) \log_2 (1-p\epsilon) 
	+ p\epsilon \log_2 \epsilon + p(1-\epsilon) \log_2 (1-\epsilon)
    } \nonumber\\
    &= p \log_2 \tup{\frac{1-p\epsilon}{p-p\epsilon}}, 
\end{align}
which is always positive. Hence, the state $\argmin_s \epsilon_s$
minimizes the mutual information
$I(X; Y_{k,s})$,
for all \(k \in \set{1,2}\) and \(P_X \in \cP(\cX)\), i.e. \(p \in [0,1]\).

Suppose now that \(\epsilon_1 = \zChannelStateA\) and \(\epsilon_2 = \zChannelStateB\).
Then, \(\epsilon = \epsilon_1 = 0.1\) minimizes the mutual informations,
and for this state, \(p_{\zChannelStateA} \approx \pgfmathparse{zChannelPopt(\zChannelStateA)}\pgfmathresult\) maximizes them.
Since different input distributions are optimal for \(W_1(\epsilon)\) and \(W_2(\epsilon)\),
the capacity region of \(\cB\) is a \emph{strict} subset of the
capacity region for parallel channels \(\cW_1, \cW_1\).
To see this, consider the input parameter \(p = p_{\zChannelStateA}\),
i.e. \(X \sim P_X = (1-p_{\zChannelStateA}, p_{\zChannelStateA})\).
Let \(R(p) = \min_{s \in \cS} \tup{ H_2(p \epsilon_s) - p H_2(\epsilon_s) }\).
The upper bound on the first rate,
\begin{equation}
  R_1 \leq \min_{s \in \cS} I(X; Y_{1,s})
      = R(p_{\zChannelStateA})
      \approx \pgfmathparse{IzChannel(zChannelPopt(\zChannelStateA), \zChannelStateA)}%
	        \pgfmathprintnumber{\pgfmathresult},
\end{equation}
is maximized by this choice of \(p\), but the upper bound on the second rate,
\begin{equation}
  R_2 \leq I(X; Y_{2,s})
      = R(1-p_{\zChannelStateA})
      \approx \pgfmathparse{IzChannel(1-zChannelPopt(\zChannelStateA), \zChannelStateA)}%
	\pgfmathprintnumber{\pgfmathresult},
\end{equation}
is suboptimal for \(\cW_2\), because
\begin{align}
  \max_{P_X \in \cP(\cX)} \min_{s \in \set{1,2}} I(X; Y_{2,s})
  &= \max_{p \in [0,1]} R(1-p) 
  = R(p_{0.1}) 
  \approx \pgfmathparse{IzChannel(zChannelPopt(\zChannelStateA),\zChannelStateA)}%
	\pgfmathprintnumber{\pgfmathresult}.
\end{align}
Therefore, we have that
\begin{align}
  \capID(\cB) = \scrR(\cB)
    &= \bigcup_{p \in [0,1]} \set{
	\begin{array}{l l}
	  (R_1, R_2) : & R_1 \leq R(p), \\
		       & R_2 \leq R(1-p)
	\end{array}
      } \nonumber\\
    &\subset \set{
	\begin{array}{l l}
	  (R_1, R_2) : & R_1 \leq \max_p R(p), \\
		       & R_2 \leq \max_p R(1-p)
	\end{array}
      } \nonumber\\
    &= \set{
	\begin{array}{l l}
	  (R_1, R_2) : & R_1 \leq \capID(\cW_1) \\
		       & R_2 \leq \capID(\cW_1)
	\end{array}
      }.
\end{align}
The capacity region \(\capSAID(\cB)\) and the square \([0, \capID(\cW_1)]^2\)
are visualized in Figure \ref{fig:zBC.capRegion},
for parameters \(\epsilon_1 = \zChannelStateA\) and \(\epsilon_2 = \zChannelStateB\).
Clearly, \(\capSAID(\cB)\) is not rectangular.


\subsection{Scalar Gaussian Channel}
\label{sec:examples.gaussian}

\begin{figure}
\begin{minipage}[b]{.49\textwidth}
  \centering
  \newif\ifqtikz\qtikzfalse


\ifqtikz

\def\capID{\mathsf{C}_{\textrm{ID}}}
\def\capSAID{\mathsf{C}_{\textrm{ID}}}
\def\capT{\mathsf{C}_{\textrm{T}}}

\usetikzlibrary{fpu}
\usetikzlibrary{math}

\input{infoth.tikz}

\edef\colorT{\iftumcd TUMRed\else red\fi}
\edef\colorID{\iftumcd TUMBlue\else blue\fi}
\edef\colorIDo{\iftumcd TUMRed\else red\fi}
\fi

\newcommand{\rateRegionGaussBC}[3][20]{

\pgfkeys{/pgf/fpu}
\pgfkeys{/pgf/fpu/output format=fixed}
\tikzmath{
 real \dda, \ddb;
 \SNRa = #2;
 \SNRb = #3;
 function R2(\x) { return Igauss((1-\x)*\SNRb/(\x*\SNRb + 1)); };
 function R1(\x) { return Igauss(\x*\SNRa); };
 \Ra = R1(1);
 \Rb = R2(0);
}
\pgfkeys{/pgf/fpu=false}

\begin{tikzpicture}[auto,inner sep=1ex, thick, node distance=2cm and 2cm, box/.style={draw,inner sep=1ex},scale=#1]
\pgfkeys{/pgf/number format=fixed}
\pgfkeys{/pgf/number format/precision=3}
\draw[->] (0,0) -- (0, {\Rb+0.15}) node[above] {$R_2$};
\draw[->] (0,0) -- ({\Ra+0.15}, 0) node[right] {$R_1$};

\draw[color=\colorID] (0, \Rb) -- +(\Ra, 0) node[below left] {$\capSAID$} -- (\Ra, 0);
\draw (0, \Rb) -- +(-0.08,0) node [left] {$\pgfmathprintnumber{\Rb}$}; 
\draw (\Ra, 0) -- +(0,-0.08) node [below] {$\pgfmathprintnumber{\Ra}$}; 

\pgfkeys{/pgf/fpu}
\pgfkeys{/pgf/fpu/output format=fixed}
\draw[dashed,color=\colorT] plot [parametric,domain=0:1,range=0:1,samples=100] ({R1(\x)}, {R2(\x)});
\pgfkeys{/pgf/fpu=false}
\draw[color=\colorT] ({R1(0.18)}, {R2(0.18)}) node[left] {$\capT$};
\end{tikzpicture}
}

\ifqtikz
\rateRegionGaussBC[2]{10^3}{10^2}
\fi
  \def\snrA{10^2}
  \def\snrB{10^1}
  \rateRegionGaussBC[2]{\snrA}{\snrB}
  \vspace{-1.2em}
  \caption[ID and transmission capacity regions of the compound scalar Gaussian BC]{%
     \label{fig:gaussian.capRegion}The ID capacity region $\capSAID$ and the
     transmission capacity region $\capT$ of the
     compound scalar Gaussian BC with $\min_s G_{1,s}^2 P = \snrA$
     and $\min_s G_{2,s}^2 P = \snrB$.}
\end{minipage}
\hfill
\begin{minipage}[b]{.49\textwidth}
  \centering
  \newif\ifqtikz\qtikzfalse


\ifqtikz

\def\capID{\mathsf{C}_{\textrm{ID}}}
\def\capSAID{\mathsf{C}_{\textrm{ID}}}
\def\capT{\mathsf{C}_{\textrm{T}}}

\usetikzlibrary{math}
\usetikzlibrary{fpu}

\input{infoth.tikz}

\edef\colorT{\iftumcd TUMRed\else red\fi}
\edef\colorID{\iftumcd TUMBlue\else blue\fi}
\edef\colorIDo{\iftumcd TUMRed\else red\fi}

\fi

\newcommand{\rateRegionMimoFive}[1][10]{

\def\ticklen{0.01}
\def\xmin{1.3}
\def\ymin{1.3}
\def\Ra{1.6711033571450173}
\def\Rb{1.5061649395916907}

\begin{tikzpicture}[auto,inner sep=1ex, thick, node distance=2cm and 2cm, scale=#1]
\pgfkeys{/pgf/number format=fixed}
\pgfkeys{/pgf/number format/precision=3}
\draw[->] ({\xmin - 0.1}, {\ymin-0.05}) -- ({\xmin - 0.1}, {\Rb + 2*\ticklen}) node[above] {$R_2$};
\draw[->] ({\xmin - 0.05}, {\ymin - 0.1}) -- ({\Ra + 2*\ticklen}, {\ymin - 0.1}) node[right] {$R_1$};

\draw (\xmin, {\ymin - 0.1}) -- +(0, -\ticklen) node[below] {$\pgfmathprintnumber{\xmin}$};
\draw (\Ra, {\ymin - 0.1}) -- +(0, -\ticklen) node[below] {$\pgfmathprintnumber{\Ra}$};
\draw ({\xmin - 0.1}, \ymin) -- +(-\ticklen, 0) node[left] {$\pgfmathprintnumber{\ymin}$};
\draw ({\xmin - 0.1}, \Rb) -- +(-\ticklen, 0) node[left] {$\pgfmathprintnumber{\Rb}$};

\pgfkeys{/pgf/fpu}
\pgfkeys{/pgf/fpu/output format=fixed}
\draw[color=\colorIDo,smooth,dashed]
  plot file {fig/rateRegion_mimo5.convex.table};
\draw[color=\colorID,smooth]
  ({\xmin-0.1}, \Rb)
  -- plot file {fig/rateRegion_mimo5.opt.table}
  -- (\Ra, {\ymin-0.1});
\pgfkeys{/pgf/fpu=false}
\draw[color=\colorID] (\Rb, \Rb) node[below left] {$\capSAID$};
\end{tikzpicture}
}

\ifqtikz
\rateRegionMimoFive
\fi
  \rateRegionMimoFive[11.1]
  \caption{\label{fig:rateRegion_mimo5} The ID capacity region
  $\capSAID$ of the compound Gaussian Product BC with
  \(G_{1,s} = \diag(1, 3, 9, 2, 3)^{-1/2}\) and
  \(G_{2,s} = \diag(3, 3, 5, 6, 1)^{-1/2}\). The dashed line
  encloses the region achieved with convex combinations
  of the two waterfillings optimal for each user.}
\end{minipage}
\end{figure}

For a scalar Gaussian BC \(\cB\), where $G_{k,s} \in \RR$,
and input power constraint $\frac{1}{n} \sum_{t=1}^n \abs{X}^2 \leq P$,
we obtain from Theorem \ref{thm:capSAID-CBC.MIMO}
the closed-form capacity formula,
\begin{equation}
        \capSAID(\cB, P)
    = \set{
    \begin{array}{l l}
      (R_1, R_2) : &\text{For all $k \in \set{1,2}$: } \\
		   &\displaystyle R_k \leq \min_{s\in\cS} 
	\frac{1}{2} \log (G_{k,s}^2 P + 1)
    \end{array}
    }.
\end{equation}
In \cite{coverBC}, the transmission capacity region of the worst scalar Gaussian
BC in $\cB$ with this input constraint is given by
\begin{align}
    \capT(\cB,P)
    &= \bigcup_{\alpha \in [0,1]} \set{
    \begin{array}{l l}
      (R_1, R_2) : &\displaystyle R_1 \leq \min_{s\in\cS} \frac{1}{2}\log(\alpha G_{1,s}^2 P + 1), \\
		   &\displaystyle R_2 \leq \min_{s\in\cS}
		   \frac{1}{2} \log \tup{\frac{(1-\alpha) G_{2,s}^2 P}{\alpha G_{2,s}^2 P + 1} }
    \end{array}
    }.
\end{align}
Figure \ref{fig:gaussian.capRegion} demonstrates that
the rectangular ID capacity region is
strictly larger that the transmission capacity region, i.e.
$\capT(\cB,P) \subset \capID(\cB,P)$.

\subsection{Gaussian Product Channel}
\label{sec:examples.MIMO}

We consider now a Gaussian product channel with \(\tau = \rho_{k,s}\),
i.e. $G_{k,s}$ are diagonal matrices, for $k \in \set{1,2}$
and $s \in \cS$.
Recall from \eqref{eq:def.R_P} and \eqref{eq:parallelMIMO}
that for every compound MIMO Gaussian channel
$\cB' = \set{B(G'_{1,s}, G'_{2,s})}_{s \in \cS}$,
we get the Gaussian product channel
$\cB = \set{B(G_{1,s}, G_{2,s})}_{s \in \cS}$
with the same capacity region
from the eigendecomposition
$G_{k,s}'^\transp G_{k,s}' = U_{k,s} G_{k,s}^2 U_{k,s}^\transp$.
Hence, this example yields the capacity region for all channels
with eigenvalue matrices $G_{k,s}^2$.

For each marginal channel in separate, 
the optimal rate  under a power constraint
$\frac{1}{n} \sum_{t=1}^n \bX_t^\transp \bX_t \leq P$ is achieved
by water-filling the transmit powers $P_1, \dots, P_\tau$
(see  \cite{telatar1999mimo}).
However, due to the different optimal water-fillings for the marginal
channels, the capacity region is not rectangular, as illustrated in  
Figure \ref{fig:rateRegion_mimo5}.

\section{Achievability Proof of Theorem \ref{thm:capSAID-CBC}}
\label{sec:proof.capSAID-CBC.achiev}

In this section we lower-bound the capacity region in Theorem \ref{thm:capSAID-CBC}. Specifically, we show that
\begin{equation}
  \capSAID(\cB, \Gamma) \supseteq \scrR_\Gamma(\cB).
\end{equation}

In the achievability proof for the DMBC, Bracher and Lapdidoth
\cite{bracherLapidoth2017idbc_arxiv,bracher2016PhD} 
first generate a single-user random code, based on a
pool-selection technique, as shown below. Then, a similar
pool-selection code is constructed for the BC using a
pair of single-user codes, one for each receiver. It is shown in
\cite{bracherLapidoth2017idbc_arxiv,bracher2016PhD} that the
corresponding ID error probabilities for the BC can be
approximated in terms of the error probabilities of the single-user
codes. We use a similar approach, and begin with the single-user
compound channel.

\subsection{Single-User Compound Channels}
\label{sec:proof.capSAID-CBC.achiev.single}

Consider the single-user compound channel \(\cW\).
We construct an ID-code for the single-user compound channel
\(\cW = \set{W_s}_{s \in \scrS}\), \(W_s : \cX \to \cP(\cY)\),
by extending the methods of Bracher and Lapidoth
\cite{bracherLapidoth2017idbc_arxiv,bracher2016PhD}
to the compound setting.
Later, we will use the results of the derivation in this section
to analyze the CBC.

\subsubsection{Code Construction}

Let \(N = \ceil{\exp e^{nR}}\) be the code size.
We fix some $\epsilon > 0$,
a PMF $P_X$ over \(\cX\) that satisfies $(1+\epsilon)\expect [\gamma(X)] \leq \Gamma$,
a pool rate \(\Rpool\), and a binning rate \(\Rbin\), such that
\begin{align}
  R &< \Rbin < \min_{s \in \cS} I(X,Y_s) \\
   \Rpool &> \Rbin
\end{align}
and
\begin{equation}
  \epsilon < \frac{\min_{s' \in \cS} I(X; Y_{s'}) - \Rbin}{2 H(X)}
  \label{eq:decEpsilon}
\end{equation}

For every index \(v \in \cV = [e^{n\Rpool}]\), perform the following.
Choose a code word \(\Xpool(v) \sim P_X^n\) at random. Then, for every
\(i \in [N]\), decide whether to add \(v\) to the set \(\bcV_i\) by a binary
experiment, with probability
\(e^{-n\Rbin}/\abs{\cV} = e^{-n(\Rpool - \Rbin)}\). That is, decide that
\(v\) is included in \(\bcV_i\) with probability \(e^{-n(\Rpool - \Rbin)}\),
and not to include with probability \(1-e^{-n(\Rpool - \Rbin)}\). Reveal
this construction to all parties.
Denote the collection of codewords and index bins by
\begin{equation}
  \CodeBL = \tup[\Big]{\{F(v)\}_{v\in\cV}, \set{\bcV_i}_{i=1}^N} \,.
\end{equation}

\subsubsection{Encoding}

To send an ID Message \(i \in [N]\), the sender chooses an index \(V\)
uniformly at random from \(\bcV_i\) and transmits the sequence \(F(V)\),
if \(\bcV_i\) is non-empty.
Otherwise, the encoder transmits $F(v^\star)$,
where $v^\star \in \cV$ is any default index.
Therefore, the encoding distribution \(\Enc[i]\) is given as follows,
\begin{equation}
  \Enc[i](x^n) = \begin{cases}
    \frac{1}{\abs{\bcV_i}} \sum_{v \in \bcV_i} \tup{
      \ind{x^n=F(v)}
    } & \text{if $\bcV_i \neq \emptyset$}, \\
      \ind{x^n = F(v^\star)} & \text{if $\bcV_i = \emptyset$}.
  \end{cases}
  \label{eq:capSAID.proof.enc.single-user}
\end{equation}

\subsubsection{Decoding}

Define the decoding region
\begin{align}
  \bcD_{i'} &= \bigcup_{v \in \bcV_{i'}} \bigcup_{s' \in \cS} \cT_\epsilon^n(P_X \times W_{s'}|F(v))
  \nonumber\\
  &=\set{ y^n\in\cY^n \,:\;
  (F(v),y^n)\in \cT_\epsilon^n(P_X \times W_{s'})\; \text{for some $v \in \bcV_{i'}$ and $s'\in\cS$}
  }.
  \label{eq:BL-code.def.decFunction}
\end{align}
The receiver receives a sequence \(Y^n\), and he is interested in a
particular message \(i'\). If \(Y^n \in \bcD_{i'}\), the receiver declares
that~\(i'\) was sent. Otherwise, he declares that~\(i'\) was not sent.

We note that our encoder is the same as in the original construction of
Bracher and Lapidoth
\cite{bracherLapidoth2017idbc_arxiv,bracher2016PhD}, for the discrete
memoryless BC. The difference between the constructions,
here and in \cite{bracherLapidoth2017idbc_arxiv,bracher2016PhD},
is in the definition of the decoder.

The ID code that is associated with the construction above is denoted by
\(\codet_\CodeBL = \set{\tup{\Enc[i], \bcD_i}}_{i=1}^N\).

\subsubsection{Error Analysis}

We will show that there exists \(\tau>0\) such that the error
probabilities of the random code \(\cC_{\CodeBL}\) satisfy
\begin{equation}
  \lim_{n\to\infty} 
  \Pr\set{
    \max_{s \in \cS} \max_{i' \in [N]} \max_{i \neq i'}
    \max\set[\big]{
      \err_0(n, \codet_{\CodeBL}, i'),\,
      \erra(W_s, n, \codet_{\CodeBL}, i'),\,
      \errb(W_s, n, \codet_{\CodeBL}, i', i)
    } \geq e^{-n\tau}
  } = 0.
  \label{eq:capSAID-CBC.direct.single-user}
\end{equation}

Let \(s\) denote the \emph{actual} channel state. As mentioned above, the
codebook that is used here is the same as
\cite{bracherLapidoth2017idbc_arxiv,bracher2016PhD}. Therefore, we can
use the cardinality bounds for the index bins \(\set{\bcV_i}\)
that were established in
\cite{bracherLapidoth2017idbc_arxiv,bracher2016PhD}.
Denote the collection of index bins by
\(\bcL= \set{\bcV_i}_{i\in [N]}\).
\begin{lemma}[see {\cite[Lemma 5]{bracherLapidoth2017idbc_arxiv}}]
Given \(\mu > 0\), let \(\cG_\mu\) be the set of all realizations \(\cL\) of
\(\bcL\) such that
\begin{align}
  \abs{\cV_i} &> (1-\delta_n)e^{n\tR}, \label{eq:Vi_lb} \\
  \abs{\cV_i} &< (1+\delta_n)e^{n\tR}, \label{eq:Vi_ub} \\
  \abs{\cV_i \cap \cV_{i'}} &< 2 \delta_n e^{n\tR} \label{eq:Vii_ub}
\end{align}
for all \(\cV_i, \cV_{i'} \in \cL, i \neq i'\), where
\(\delta_n = e^{-n\mu/2}\). Then, the probability that \(\bcL \in \cG_\mu\)
converges to 1 as \(n\to\infty\), i.e.
\begin{equation}\label{eq:VinG}
  \lim_{n\to\infty} \Pr\set{\bcV_\star \in \cG_\mu} = 1,
\end{equation}
for \(\mu < \Rpool - \Rbin\).
\label{lemma:BL-binning.cardinalities}
\end{lemma}
Hence, it suffices to consider the bin collection realizations \(\cL\) of
\(\bcL\) that satisfy \eqref{eq:Vi_ub}--\eqref{eq:Vii_ub},
for \(\mu \in (0, \Rpool - \Rbin)\).
Thus, the encoding distribution is
given by
\begin{equation}
  \Enc[i](x^n) = \frac{1}{|\cV_i|} \sum_{v \in \cV_i} \ind{x^n = F(v)},
\end{equation}
for \(i \in [N]\), due to
\eqref{eq:capSAID.proof.enc.single-user}
and \(\cV_i \neq \emptyset\) by \eqref{eq:Vi_lb}.

\paragraph{Encoding Error}

First, we bound the probability of an encoding error, i.e.
\begin{equation}
  \err_0(n, \codet, i)
    = \sum_{x^n \in \cX} \enc[i][x^n] \ind{\frac{1}{n} \sum_{t=1}^n \gamma(X_t) > \Gamma}
    = \frac{1}{\abs{\cV_i}} \sum_{v \in \cV_i} \ind[\Big]{ \sum_{t=1}^n \gamma\tup{F(v)_t} > n\Gamma },
\end{equation}
where $F(v)_t$ is the $t$-th letter of $F(v)$.
For every $x^n \in \cT_\epsilon^n(P_X)$ and sufficiently large $n$, the input constraint is satisfied, since
\begin{equation*}
  \sum_{t=1}^n \gamma\tup{x_t} \leq n (1+\epsilon) \expect [\gamma\tup{X}] \leq n \Gamma,
\end{equation*}
by the Typical Average Lemma \cite{elgamalKim2011network_it}.
Therefore,
\begin{align}
  \expect_\CodeBL\intv[\Big]{ \ind[\Big]{ \sum_{t=1}^n \gamma\tup{F(v)_t} > n \Gamma } }
  &= \sum_{x^n \in \cX^n} P_X^n(x^n) \ind[\Big]{ \frac{1}{n} \sum_{t=1}^n \gamma\tup{x_t} > \Gamma } \nonumber\\ &
  \leq \Pr\set{ X^n \notin \cT_\epsilon^n(P_X) } \nonumber\\ &
  \leq e^{-n\tau_0},
\end{align}
for some $\tau_0 > 0$, where the last inequality follows from Lemma \ref{lemma:X_is_untypical_ub}.
Now, we show that the encoding error probability is small with high probability.
Let \(\alpha\) be such that
\begin{equation}
  0 < \alpha < (\Rbin - R)/2.
  \label{eq:alphaRbin}
\end{equation}
Then, by the union bound,
\begin{align}
  \Pr
  &\set{
    \max_{i \in [N]}
    \err_0(n, \codet_{\CodeBL}, i)
    \geq e^{-n\tau_0} + e^{-n\alpha}
    } \nonumber\\
  &= \Pr\set{
    \exists\,i \in [N] :
    \frac{1}{\abs{\cV_{i}}} \sum_{v \in \cV_{i}} \ind[\Big]{ \sum_{t=1}^n \gamma(F(v)_t) > n \Gamma }
    \geq e^{-n\tau_0} + e^{-n\alpha}
    } \nonumber\\
  &\leq \sum_{i \in [N]}
    \Pr\set{
    \frac{1}{\abs{\cV_{i}}} \sum_{v \in \cV_{i}} \ind[\Big]{ \sum_{t=1}^n \gamma(F(v)_t) > n \Gamma }
    \geq e^{-n\tau_0} + e^{-n\alpha}
    }.
\end{align}
By \eqref{eq:Vi_lb},
\(\abs{\cV_{i}} > e^{n\Rbin}/2\), for all $i \in [N]$ and sufficiently large $n$. By symmetry,
since the code words \(F(v), v \in \cV\) are i.i.d., the probability terms
are bounded by
\(\exp \tup{ -2 e^{-2n\alpha} \abs{\cV_{i}} }\leq \exp \tup{ -e^{-2n\alpha}e^{n\Rbin} }\),
by Hoeffding's inequality (Theorem \ref{thm:hoeffding}).
Hence, since \(N \leq \exp e^{nR}\), there exists \(\theta_0 > 0\) such that
\begin{align}
  \Pr &\set{
    \max_{i \in [N]}
    \err_0(n, \codet_{\CodeBL}, i)
    \geq e^{-n\theta_0}
  } \nonumber\\
  &\leq N \exp\tup{e^{-2n\alpha} e^{n\Rbin}} \nonumber\\
  &\leq \exp\tup{e^{nR} - e^{-2n\alpha} e^{n\Rbin}}
  \label{eq:ranCode.error.idEnc.proof}
\end{align}
which tends to zero as \(n\to\infty\), by \eqref{eq:alphaRbin}.

\paragraph{Missed ID Error}

Next, we bound the probability of the missed-ID error (first kind), given by
\begin{align}
  \erra(W_s, n, \codet_\CodeBL, i)
    &= \sum_{x^n \in \cX^n} \Enc[i](x^n) W^n_s\tup{\bcD_{i}^c \middle| x^n} 
    = \frac{1}{\abs{\cV_{i}}} \sum_{v \in \cV_{i}} W^n_s\tup{\bcD_{i}^c \middle| F(v)}.
    \label{eq:capSAID-CBC.missed-ID.expression}
\end{align}

Consider any index bin \(\cV_i \in \cL\). Let
\((X^n, Y^n_s) \sim P_{XY_s}^n = P_X^n \times W_s^n\),
and recall that by \eqref{eq:BL-code.def.decFunction},
\begin{align*}
  \bcD_{i} &= \bigcup_{v' \in \cV_{i}} \bigcup_{s' \in \cS} \cT_\epsilon^n(P_{XY_{s'}}|\Xpool(v')).
\end{align*}
Therefore,
\begin{align}
    W^n_s \tup{\bcD_i^c \middle| x^n}
     &= \Pr\tup[\Big]{
       Y^n_s \notin \bigcup_{s' \in \cS} \bigcup_{v' \in \cV_i}
       \cT_\epsilon^n\tup{P_{XY_{s'}}|F(v')}| X^n=x^n
     }.
     \label{eq:capSAID-CBC.missed-ID.givenXn}
\end{align}
Observe that in general, the event
\(Y^n_s \notin \bigcup_{s'\in\cS,v'\in\cV_i} \cA(s',v')\) implies that
\(Y^n_s \notin  \cA(s,v)\), for $v \in \cV_i$.
Therefore, we have
\begin{equation}
  W^n_s \tup{\bcD_i^c \middle| x^n}
    \leq \Pr\tup[\Big]{ Y^n_s \notin \cT_\epsilon^n\tup{P_{XY_s}|F(v)}| X^n=x^n }.
\end{equation}
Averaging over the realizations of \(F(v)\), we obtain
\begin{align}
  \expect_\CodeBL \intv{W^n_s \tup{\bcD_i^c \middle| \Xpool(v)}}
    &= \sum_{x^n\in\cX^n} P_X^n(x^n)W^n_s \tup{\bcD_i^c \middle| x^n} \nonumber\\
    &\leq \Pr\tup[\big]{ (X^n,Y^n_s) \notin \cT^n_\epsilon\tup{P_{XY_s}} }, \nonumber\\
    &< e^{-n\tau_1},
\end{align}
for some \(\tau_1 > 0\), where the
last inequality follows from Lemma
\ref{lemma:X_is_untypical_ub}.
Now, we show that the missed-ID error is small with high probability.
As in the proof for the encoding error, it follows from the union bound that
\begin{align}
  \Pr
  &\set{
    \max_{s \in \cS} \max_{i \in [N]}
    \erra(W_s, n, \codet_{\CodeBL}, i)
    \geq e^{-n\tau_1} + e^{-n\alpha}
    } \nonumber\\
  &\leq \sum_{s \in \cS} \sum_{i \in [N]}
    \Pr\set{
    \frac{1}{\abs{\cV_{i}}} \sum_{v \in \cV_{i}} W^n_s\tup{ \bcD_{i}^c \middle| F(v) }
    \geq e^{-n\tau_1} + e^{-n\alpha}
    }.
\end{align}
Hence, by symmetry and Hoeffding's inequality (Theorem \ref{thm:hoeffding}),
the probability terms are bounded by
\(\exp \tup{ -2 e^{-2n\alpha} \abs{\cV_{i}} }\leq \exp \tup[\big]{ -e^{-2n\alpha}e^{n\Rbin} }\),
for sufficiently large $n$.
Therefore, there exists \(\theta_1 > 0\) such that
\begin{align}
  \Pr &\set{
    \max_{s \in \cS} \max_{i \in [N]}
    \erra(W_s, n, \codet_{\CodeBL}, i)
    \geq e^{-n\theta_1}
  } \nonumber\\
  &\leq \abs\cS N \exp\tup{e^{-2n\alpha} e^{n\Rbin}} \nonumber\\
  &\leq \abs\cS \exp\tup{e^{nR} - e^{-2n\alpha} e^{n\Rbin}}
  \label{eq:ranCode.error.idOne.proof}
\end{align}
which tends to zero as \(n\to\infty\) by \eqref{eq:alphaRbin}.

\paragraph{False ID Error}

Last, we bound the probability of the false-ID error (second kind).
Suppose that the sender sends an ID message \(i\) and the receiver is
interested in \(i' \neq i\). Recall that we can restrict our attention to
realizations \(\cL= \set{\cV_i} \in \cG_\mu\), following Lemma
\ref{lemma:BL-binning.cardinalities}.
Observe that
\begin{align}
  \errb(W_s, n, \codet_\CodeBL, i', i)
    &= \sum_{x^n \in \cX^n} \Enc[i](x^n) W^n_s\tup{\bcD_{i'} \middle| x^n} \nonumber\\
    &= \frac{1}{\abs{\cV_i}} \sum_{v \in \cV_i} W^n_s\tup{\bcD_{i'} \middle| F(v)} \nonumber\\
    &=  \frac{1}{\abs{\cV_i}}\sum_{v \in \cV_i \cap \cV_{i'}}
       W^n_s\tup{ \bcD_{i'} \middle| F(v) }
       + \frac{1}{\abs{\cV_i}}\sum_{v \in \cV_i \cap \cV_{i'}^c}
       W^n_s\tup{ \bcD_{i'} \middle| F(v) } \nonumber\\
    &\leq \frac{1}{\abs{\cV_i}}\abs{\cV_i \cap \cV_{i'}}
    + \frac{1}{\abs{\cV_i \cap \cV_{i'}^c}}\sum_{v \in \cV_i \cap \cV_{i'}^c}
       W^n_s\tup{ \bcD_{i'} \middle| F(v) }
      \label{eq:BL-code.errB.split}
\end{align}
where the inequality
holds since \(\abs{\cV_i \cap \cV_{i'}^c}\leq \abs{\cV_i }\). By Lemma \ref{lemma:BL-binning.cardinalities},
the first term is bounded by
\begin{equation}
  \frac{\abs{\cV_i \cap \cV_{i'}}}{\abs{\cV_i}}
  < \frac{2\delta_n}{1-\delta_n}
  < \delta_n
  \label{eq:BL-binning.Vii_ub.pretty}
\end{equation}
(see \eqref{eq:Vi_lb} and \eqref{eq:Vii_ub}), where the second inequality holds as
\(\delta_n < 1/2\), for sufficiently large \(n\).

It remains to bound the second term in the right-hand side of
\eqref{eq:BL-code.errB.split}, for which
\(v\in \cV_i\cap \cV_{i'}^c\). Namely, \(v\in\cV_i\) and \(v\notin\cV_{i'}\).
Let \(X^n = F(v)\) be the transmitted codeword, hence \(Y^n_s\) is the
corresponding channel output for the actual channel state \(s\). Consider
a codeword \(F(v')\) that is tested by the receiver, where
\(v' \in \cV_{i'}\). For every pair of indices \(v \notin \cV_{i'}\) and
\(v' \in \cV_{i'}\), we have that the code word \(F(v')\) and the channel
output \(Y^n_s\) are independent, i.e.
\begin{align}
  (F(v'), Y^n_s) \sim P_{X}^n(x) \cdot P^n_{Y_s}(y). \label{eq:BL-code.errB.indep}
\end{align}

Now, we consider the expecation of
\(\errb(W_s, n, \codet_\CodeBL, i', i)\). By expanding \(\bcD_{i'}\) and
applying the union bound, we obtain
\begin{align}
  \expect_\CodeBL \intv{W^n_s \tup{\bcD_{i'} \middle| \Xpool(v)}}
    &= \Pr \set[\Big]{
	(F(v'), Y^n_s) \in \bigcup_{v' \in \cV_1} \bigcup_{s' \in \cS}
	\cT^n_\epsilon\tup{P_{X Y_{s'}}}
      } \nonumber\\
    &\leq \sum_{v' \in \cV_1} \sum_{s' \in \cS}
      \Pr \set[\Big]{
	(F(v'), Y^n_s) \in \cT^n_\epsilon\tup{P_{XY_{s'}}}
      } \nonumber\\
    &= \sum_{v' \in \cV_1} \sum_{s' \in \cS}
      \sum_{y^n \in \cY^n} P_{Y_s}^n(y^n)
      P_X^n \tup[\Big]{
	(F(v'), y^n) \in \cT^n_\epsilon\tup{P_{X Y_{s'}}}
      },
      \label{eq:BL-code.errB.expect.indep}
\end{align}
where the last equality follows from
\eqref{eq:BL-code.errB.indep}. By Lemma \ref{lemma:joint_typicality},
\begin{equation}
  P_X^n \tup[\Big]{
    (X^n, y^n) \in \cT^n_\epsilon\tup{P_{X Y_{s'}}}
  } \leq e^{-n[I(X; Y_{s'})-2\epsilon H(X)]}
  \label{eq:BL-code.errB.expect.typBound}
\end{equation}
for all \(y^n \in \cY^n\). Therefore, by
\eqref{eq:BL-code.errB.expect.indep} and
\eqref{eq:BL-code.errB.expect.typBound},
there exists \(\tau_2 > 0\) such that
\begin{align}
  \expect_\CodeBL \intv{W^n_s \tup{\bcD_1 \middle| \Xpool(v)}}
    &\leq \abs{\cS} \abs{\cV_1} \max_{s' \in \cS} e^{-n[I(X; Y_{s'})-2\epsilon H(X)]} \nonumber\\
    &\overset{(a)}{<} \abs\cS (1+\delta_n) e^{-n(\min_{s' \in \cS} I(X; Y_{s'}) - 2\epsilon H(X) - \Rbin)} \nonumber\\
    &\overset{(b)}{<} e^{-n\tau_2},
    \label{eq:BL-code.error.expectTwo}
\end{align}
where (a) holds since by \eqref{eq:Vi_lb}, \(\abs{\cV_1} < (1+\delta_n)e^{n\Rbin}\),
and (b) holds since by \eqref{eq:decEpsilon}, \(\min_{s' \in \cS} I(X; Y_{s'}) - 2\epsilon H(X) - \Rbin > 0\).
We show now that the false-ID error is small with high probability.
By the union bound,
\begin{align}
  \Pr &\set{
   \max_{s \in \cS} \max_{i' \in [N]} \max_{i \neq i'}
  \errb(W_s, n, \codet_{\CodeBL}, i', i)
       \geq \delta_n + e^{-n\tau_2} + e^{-n\alpha}
  }\nonumber\\
  &= \Pr \set{
   \exists\, s \in \cS, i, i' \in [N] , i \neq i' :
   \frac{1}{\abs{\cV_i}} \sum_{v \in \cV_i}
       W^n_s\tup{ \bcD_{i'} \middle| F(v) }
       \geq \delta_n + e^{-n\tau_2} + e^{-n\alpha}
  } \nonumber\\
  &\leq \sum_{s \in \cS} \sum_{i' \in [N]} \sum_{i \neq i'}
    \Pr\set{
    \frac{1}{\abs{\cV_i}} \sum_{v \in \cV_i}
       W^n_s\tup{ \bcD_{i'} \middle| F(v) }
    \geq \delta_n + e^{-n\tau_2} + e^{-n\alpha}
    }.
    \label{eq:BL-code.errB.boundA}
\end{align}
Note that by \eqref{eq:BL-code.errB.split} and \eqref{eq:BL-binning.Vii_ub.pretty},
\begin{equation}
   \frac{1}{\abs{\cV_i}} \sum_{v \in \cV_i} W^n_s\tup{ \bcD_{i'} \middle| F(v) }
     \leq \delta_n + \frac{1}{\abs{\cV_i \cap \cV_{i'}^c}}\sum_{v \in \cV_i \cap \cV_{i'}^c}
       W^n_s\tup{ \bcD_{i'} \middle| F(v) }.
\end{equation}
Therefore there exists \(\theta_2 > 0\) such that
\begin{align}
  \Pr
  &\set{
   \max_{s \in \cS} \max_{i' \in [N]} \max_{i \neq i'}
   \errb(W_s, n, \codet_{\CodeBL}, i', i)
       \geq \delta_n + e^{-n\theta_2}
  } \nonumber\\
  &\leq \sum_{s \in \cS} \sum_{i' \in [N]} \sum_{i \neq i'}
    \Pr\set{
    \frac{1}{\abs{\cV_i \cap \cV_{i'}^c}} \sum_{v \in \cV_i \cap \cV_{i'}^c}
       W^n_s\tup{ \bcD_{i'} \middle| F(v) }
    \geq e^{-n\tau_2} + e^{-n\alpha}
    } \nonumber\\
  &\overset{(a)}{\leq} \abs{\cS} N^2 \exp\tup{-2 e^{-2n\alpha} \abs{\cV_i \cap \cV_{i'}^c}} \nonumber\\
  &\overset{(b)}{<} \abs\cS \exp\tup{2e^{nR} - e^{-2n\alpha} e^{-n\Rbin}} 
  , \label{eq:ranCode.error.idTwo.proof}
\end{align}
sufficiently large \(n\), where (a) follows from
Hoeffding's inequality (Therorem \ref{thm:hoeffding}),
since the codewords \(F(v), v \in \cV\) are i.i.d., and (b) follows from
\(N \leq \exp \tup{e^{nR}}\), and
\begin{equation}
  \abs{\cV_i \cap \cV_{i'}^c}
  = \abs{\cV_i} - \abs{\cV_i \cap \cV_{i'}}
  > (1-\delta_n) e^{n\Rbin} - 2\delta_n e^{n\Rbin}
  \geq e^{n\Rbin}/2,
\end{equation}
as \(\abs{\cV_i} \geq (1-\delta_n) e^{n\Rbin}\) and
\(\abs{\cV_i \cap \cV_{i'}} < 2\delta_n e^{n\Rbin}\), by Lemma \ref{lemma:BL-binning.cardinalities}
(see \eqref{eq:Vi_lb} and \eqref{eq:Vii_ub}, respectively),
where the last inequality follows from \(\delta_n < 1/2\),
for sufficiently large \(n\).

Based on \eqref{eq:ranCode.error.idEnc.proof},
\eqref{eq:ranCode.error.idOne.proof}
and \eqref{eq:ranCode.error.idTwo.proof}, we
have established that \eqref{eq:capSAID-CBC.direct.single-user}
holds for \(\tau = \min\set{\theta_0, \theta_1,\theta_2}\).

\subsection{Broadcast Channels}
\label{sec:proof.capSAID-CBC.achiev.broadcast}

In this section, we show the direct part for the ID capacity
region of the CBC. That is, we show that \(\capSAID(\cB) \supseteq \scrR(\cB)\).
The analysis makes use of the our single-user derivation above.

\subsubsection{Code Construction}

We extend Bracher and Lapdioth's \cite{bracherLapidoth2017idbc_arxiv,bracher2016PhD}
idea to combine two BL code books \(\CodeBL_1, \CodeBL_2\) that share the same pool.
Fix a PMF \(P_X\) over \(\cX\) and rates \(R_k,\Rbin_k\), for \(k\in\{1,2\}\), that satisfy
\begin{subequations} \label{eq:capSAID.CBC.achiev.constr.rates}
\begin{IEEEeqnarray}{rcL}
  R_1 < &\Rbin_1        &< \min_{s \in \cS} I(X; Y_{1,s}) \\
  R_2 < &\Rbin_2        &< \min_{s \in \cS} I(X; Y_{2,s}) \\
	& \max\set[\big]{\Rbin_1, \Rbin_2} &< \Rpool \\
  \Rpool <& \Rbin_1 + \Rbin_2.
\end{IEEEeqnarray}
\end{subequations}
Let $N_k = e^{nR_k}$.
For every index \(v \in \cV = [e^{n\Rpool}]\), perform the following. Choose
a code word \(F(v) \sim P^n_X\) at random, as in the single-user case.
Then, for every \(i_k\), decide whether to add \(v\) to the set \(\bcV_{k,i_k}\) by a binary
experiment, with probability
\(e^{-n\Rbin_k}/\abs{\cV} = e^{-n(\Rpool - \Rbin_k)}\). That is, decide that
\(v\) is included in \(\bcV_{k,i_k}\) with probability \(e^{-n(\Rpool - \Rbin_k)}\),
and not to include with probability \(1-e^{-n(\Rpool - \Rbin_k)}\). 
Finally, for every pair \((i_1,i_2) \in [N_1] \times [N_2]\), select a common index \(V_{i_1,i_2}\)
uniformly at random from \(\bcV_{1,i_1} \cap \bcV_{2,i_2}\), if this intersection is non-empty.
Otherwise, if \(\bcV_{1,i_1} \cap \bcV_{2,i_2}=\emptyset\), then draw \(V_{i_1,i_2}\) uniformly from \(\cV\).
Reveal this construction to all parties.

Denote the collection of codewords and index bins for the compound BC by
\begin{equation}
  \CodeBL_B = \tup{\Xpool
    , \set{\bcV_{1,i_1}}_{i_1 \in [N_1]}, \set{\bcV_{2,i_2}}_{i_2 \in [N_2]}
    , \set{V_{i_1,i_2}}_{(i_1,i_2) \in [N_1] \times [N_2]}
    }.
\end{equation}
Note that, for $k \in \set{1,2}$, \(\CodeBL_B\) includes all elements of
\begin{equation*}
  \CodeBL_k = \tup{ \Xpool, \set{\bcV_{k,i_k}}_{i_k \in [N_k]} },
\end{equation*}
defined for the single-user marginal channels $\cW_k$
as in Section \ref{sec:proof.capSAID-CBC.achiev.single}.
We denote the corresponding single-user code by
\begin{equation*}
  \codet_{\CodeBL_k} = \set[\big]{(\EncAlt[k,i_k], \bcD_{k,i_k})}_{i_k=1}^{N_k}.
\end{equation*}

\subsubsection{Encoding}

To send an ID message pair \((i_1,i_2) \in [N_1] \times [N_2]\), the
sender transmits the sequence \(F(V_{i_1,i_2})\).
Therefore, given \(\CodeBL_B\), the encoding distribution \(\Enc[i_1,i_2]\) is given
by the following \(0\)-\(1\)-rule:
\begin{equation}
  \Enc[i_1,i_2][x^n] = \ind{x^n = \Xpool(V_{i_1,i_2})},
\end{equation}
for \(x^n\in\cX^n\).
We note that for every realization \(\codet_{\codeBL_B}\) of a BL
code, this encoding function is deterministic.

\subsubsection{Decoding}

Receiver \(k\), for \(k=1,2\), employs the decoder of the
single-user code \(\codet_{\CodeBL_k}\).
Specifically, suppose that Receiver $k$ is interested in an ID message \(i'_k \in [N_k]\).
Then, he uses the decoding set \(\bcD_{k,i'_k}\) to decide whether \(i'_k\) was sent or not.

We denote the ID code for the CBC that is associated with the construction above by
\begin{equation}
  \codet_{\CodeBL_B} = \set[\big]{(\Enc[i_1,i_2], \bcD_{1,i_1}, \bcD_{2,i_2}) : i_1 \in [N_1],~i_2 \in [N_2]}.
\end{equation}

\subsubsection{Error Analysis}

Based on the idea of Bracher and Lapidoth \cite{bracherLapidoth2017idbc_arxiv,bracher2016PhD},
we will show that the semi-average error probabilities of the ID code
defined above can be approximately upper-bounded by the respective error probabilities
of the single-user ID-codes \(\codet_{\CodeBL_1}\) and \(\codet_{\CodeBL_2}\) for the
respective receivers.

Denote the \emph{total variation distance}
between two PMFs \(P\) and \(Q\) over a given finite set \(\cA\). It
is defined as
\begin{align*}
  d(P,Q)
    &= \max_{\cA' \subseteq \cA} \tup{P(\cA') - Q(\cA')} \\
    &= \frac{1}{2}\sum_{a \in \cA} \abs{P(a) - Q(a)}.
\end{align*}
For the second equality see (11.137) in \cite{coverThomas2005elements}.

We consider now only Receiver 1 and his
marginal channel \(\cW_1\).
Since the code construction is completely symmetric between the
two receivers, the same arguments hold for Receiver 2 and \(\cW_2\).
The encoding distribution of \(\codet_{\CodeBL_B}\) can be expressed as
\begin{equation*}
  \Enc[i_1,i_2][x^n] = \sum_{v \in \cV} \ind{x^n = \Xpool(v)} \bQ_{i_1,i_2}(v),
\end{equation*}
where $\bQ_{i_1,i_2}(v) = \ind{V_{i_1,i_2} = v}$.
Similarly, the encoding distribution of the single-user code
\begin{equation*}
\codet_{\CodeBL_k} = \set[\big]{(\EncAlt[k,i_k], \bcD_{k,i_k})}_{i_k \in [N_k]}
\end{equation*}
can be rewritten as
\begin{equation*}
  \EncAlt[k,i_k](x^n) = \sum_{v \in \cV}  \ind{x^n = F(v)} \tbQ_{k,i_k}(v)
\end{equation*}
for some family of PMFs \(\tbQ_{k,i_k} \in \cP(\cV)\), \(i_k \in [N_k]\).
Thus, for every BC \(B_s,\,s \in \cS\), the error probabilities are bounded by
\begin{subequations}
\label{eq:BL-code.err_reduces}
\begin{align}
  \saerr_{1,0}(n, \codet_{\CodeBL_B}, i_1)
    &= \frac{1}{N_2} \sum_{i_2 \in [N_2]} \sum_{x^n \in \cX^n} \Enc[i_1,i_2][x^n] \ind[\Big]{\sum_{t=1}^n \gamma(X_t) > n \Gamma} \nonumber\\
    &\leq \sum_{x^n \in \cX^n} \EncAlt[1,i_1][x^n] \ind[\Big]{\sum_{t=1}^n \gamma(X_t) > n \Gamma} +
      d\tup[\bigg]{
	\frac{1}{N_2} \sum_{i_2 \in [N_2]} \Enc[i_1,i_2]\,,~
	\EncAlt[1,i_1]
      } \nonumber\\
    &= \err_0(n, \codet_{\CodeBL_1}, i_1) +
      d\tup[\bigg]{
	\frac{1}{N_2} \sum_{i_2 \in [N_2]} \Enc[i_1,i_2]\,,~
	\EncAlt[1,i_1]
      },
      \label{eq:BL-code.errEnc_reduces}
      \\
  \saerrya(B_s, n, \codet_{\CodeBL_B}, i_1)
    &= \frac{1}{N_2} \sum_{i_2 \in [N_2]} \Enc[i_1,i_2] W^n_{1,s}(\bcD_{1,i_1}^c) \nonumber\\
    &\leq \EncAlt[1,i] W^n_{1,s}(\bcD_{1,i_1}^c) +
      \delta
      \nonumber\\
    &= \erra(B_s, n, \codet_{\CodeBL_1}, i_1) +
      \delta
      \label{eq:BL-code.errI_reduces}
      \\
  \saerryb(B_s, n, \codet_{\CodeBL_B}, i'_1, i_1)
    &= \frac{1}{N_2} \sum_{i_2 \in [N_2]} \Enc[i_1,i_2] W^n_{1,s}(\bcD_{1,i'_1}) \nonumber\\
    &\leq \EncAlt[1,i_1] W^n_{1,s}(\bcD_{1,i'_1}) +
      \delta
      \nonumber\\
    &= \errb(B_s, n, \codet_{\CodeBL_1}, i'_1, i_1) +
      \delta,
      \label{eq:BL-code.errII_reduces}
\end{align}
\end{subequations}
where
$
  \delta = 
    d\tup[\big]{
      \frac{1}{N_2} \sum_{i_2 \in [N_2]} \Enc[i_1,i_2] W^n_{1,s}\,,~
      \EncAlt[1,i_1] W^n_{1,s}
    }.
$
The error probabilities \(\saerr_{2,0}, \saerrza, \saerrzb\) for Receiver 2 are bounded analogously,
in terms of the error probabilities for \(\codet_{\CodeBL_2}\).
By the data-proccessing inequality for the total variation distance \cite[Lemma 1]{cannoneRonServedio2015testing_arxiv},
\begin{equation}
   \delta \leq d\tup[\bigg]{
     \frac{1}{N_2} \sum_{i_2 \in [N_2]} \bQ_{i_1,i_2}\,,~
     \tbQ_{1,i_1}
   },
   \label{eq:tvd.dataProcessing}
\end{equation}
which is independent of the channel.

Now, Let \(\tbQ_{2,i_2}\) be a single-user encoding distribution for \(\cW_2\),
analogous to \(\tbQ_{1,i_1}\).
The next lemma bounds the right-hand side
of \eqref{eq:tvd.dataProcessing}
to zero in probability as \(n \to \infty\).

\begin{lemma}[see {\cite[Equations (98, 113--156)]{bracherLapidoth2017idbc_arxiv}}]
For some \(\tau > 0\)
\begin{equation}
  \lim_{n\to\infty} \Pr\set{
    \max_{i_1 \in [N_1]} 
      d\tup[\bigg]{
	\frac{1}{N_2} \sum_{i_2 \in [N_2]} \bQ_{i_1,i_2}\,,~
	\tbQ_{1,i_1}
      }
    \geq e^{-n\tau}
  } = 0
\end{equation}
and
\begin{equation}
  \lim_{n\to\infty} \Pr\set{
    \max_{i_2 \in [N_2]} 
      d\tup[\bigg]{
	\frac{1}{N_1} \sum_{i_1 \in [N_1]} \bQ_{i_1,i_2}\,,~
	\tbQ_{2,i_2}
      }
    \geq e^{-n\tau}
  } = 0.
\end{equation}
\label{lemma:BL-binning.tvd-converges}
\end{lemma}

By this lemma and \eqref{eq:tvd.dataProcessing}, the total variation distances in
\eqref{eq:BL-code.err_reduces} are upper-bounded with high probability,
for sufficiently large \(n\).
Hence, the error probabilities for the BC code \(\codet_{\CodeBL_B}\)
are approximately upper-bounded by the corresponding error probabilities
for the single-user marginal codes \(\codet_{\CodeBL_1}, \codet_{\CodeBL_2}\).

By \eqref{eq:capSAID-CBC.direct.single-user} for the single-user
compound channel,
given any receiver $k \in \set{1,2}$, state $s \in \cS$,
and message pair \(i_k, i'_k \in [N_k]\) such that \(i_k \neq i'_k\),
the error probabilities 
  \(\err_0(n, \codet_{\CodeBL_k}, i_k)\),
  \(\erra(W_{k,s}, n, \codet_{\CodeBL_k}, i_k)\) and
  \(\errb(W_{k,s}, n, \codet_{\CodeBL_k}, i'_k, i_k)\)
converge with high probability
to zero exponentially in \(n\).
This completes the proof of the direct part.
\qed

\section{Converse Proof of Theorem \ref{thm:capSAID-CBC}}
\label{sec:proof.capSAID-CBC.converse}

Next, we upper-bound the capacity region in Theorem \ref{thm:capSAID-CBC}, i.e.
we prove that
\begin{equation}
  \capSAID(\cB, \Gamma) \subseteq \scrR_\Gamma(\cB).
\end{equation}

Bracher and Lapidoth \cite[Claim 16]{bracherLapidoth2017idbc_arxiv}
modified the single-user converse from \cite{hanVerdu1992idNewResults}
so that the bounds for the single-user marginal channels
can be combined for the DMBC.
We additionally combine the bounds
for all states of the CBC.

We denote \(I(P_X; P_{Y|X}) = I(X; Y)\).
The key lemma we use is the following.

\begin{lemma}[see Lemma 21 in {\cite{bracherLapidoth2017idbc_arxiv}}]
For every discrete memoryless channel \(W : \cX \to \cP(\cY)\),
every ID rate \(R\), all positive constants
\(\lambda, \epsilon, \kappa\) satisfying \(2\lambda < \kappa < 1\), \(N = \exp e^{nR}\) and sufficiently large \(n\),
if \(\set{(\enc[i], \cD_i)}_{i=1}^N\) is an \((N, n, \lambda)\) ID-code for \(W\), then
\begin{equation}
  \frac{1}{N} \sum_{i=1}^N \enc[i][X^n \in \set{x^n \in \cX^n : I(\type{P}_{x^n}, W) \leq R - \epsilon}]
        < \kappa - \exp\tup{-e^{nR/2}}.
\end{equation}
\label{lemma:dmc.probTypeAVG}
\end{lemma}

Consider now an \((N_1, N_2, n, \lambda)\) ID code,
\(\codet = \set{(\enc[i_1,i_2], \cD_{1,i_1}, \cD_{2,i_2}) : i_1 \in [N_1], i_2 \in [N_2]}\),
for the CBC \(\cB\), and the PMFs
\begin{align*}
        \enc[1,i_1] &= \frac{1}{N_2} \sum_{i_2=1}^{N_2} \enc[i_1,i_2], \\
        \enc[2,i_2] &= \frac{1}{N_1} \sum_{i_1=1}^{N_1} \enc[i_1,i_2], \\
        \enc[]' &= \frac{1}{N_1 N_2} \sum_{i_1=1}^{N_1} \sum_{i_2=1}^{N_2} \enc[i_1,i_2].
\end{align*}
Thus, we get single-user codes $\codet_k = \set{ \enc[k,i_ki], \cD_{k,i_k} }_{i_k=1}^{N_k}$, for $k \in \set{1,2}$.
By the definition of the error probabilities \eqref{eq:def.SAID-errProbs},
\begin{subequations}
\begin{align}
  \lambda & > \saerrya(B_s, n, \codet, i_k) \nonumber\\
    &= \sum_{x^n \in \cX^n} \frac{1}{N_2} \sum_{i_2=1}^{N_2}
     \enc[i_1,i_2](x^n) W^n_1(\cD_{1,i_1}^c|x^n), \nonumber\\
    &= \sum_{x^n \in \cX^n}
     \enc[1,i_1](x^n) W^n_{1,s}(\cD_{1,i_1}^c|x^n), \nonumber\\
     &= \erra(W_{1,s}, n, \codet_1, i_1)
\end{align}
for all \(s \in \cS\), and similarly we get
\begin{align}
  \lambda & > \saerryb(B_s, n, \codet, i'_1, i_1) = \erra(W_{1,s}, n, \codet_1, i'_1, i_2), \\
  \lambda & > \saerrza(B_s, n, \codet, i_2) = \erra(W_{2,s}, n, \codet_2, i_2), \\
  \lambda & > \saerryb(B_s, n, \codet, i'_2, i_2) = \erra(W_{2,s}, n, \codet_2, i'_2, i_2),
\end{align}
for all \(s \in \cS\).
\end{subequations}
Hence, for every $k \in \set{1,2}$ and $s \in \cS$,
$\codet_k$ is an $(N_k, n, \lambda)$ ID-code for $W_{k,s}$.
Let \(R_k = \frac{1}{n} \log\log N_k\).
Since for \(s \in \cS\) and \(k \in \set{1,2}\) the marginal channel \(W_{k,s}\) is a discrete
memoryless channel,
Lemma \ref{lemma:dmc.probTypeAVG} proves that
for all \(s \in \cS\), all constants \(\epsilon > 0\), \(\lambda \in [0,\frac{1}{4\abs\cS})\),
\(\kappa \in (2\lambda, \frac{1}{2\abs\cS})\), and sufficiently large \(n\),
\begin{equation}
  \enc[]'\tup{ X^n \in \set{x^n \in \cX^n : I(\type{P}_{x^n}, W_{k,s}) \leq R_k - \epsilon} }
        < \kappa + \exp\tup{-e^{nR/2}}.
  \label{eq:converse.probType.dmc}
\end{equation}
Let \(X^n\sim \enc[]'\). Then, by the union bound,
\begin{align}
  \Pr &\tup{ X^n \in \bigcap_{s \in \cS} \set{
    \begin{array}{l l}
      x^n \in \cX^n : & I(\type{P}_{x^n}, W_{1,s}) > R_1 - \epsilon, \\
     		      & I(\type{P}_{x^n}, W_{2,s}) > R_2 - \epsilon
    \end{array}
  } } \nonumber\\
  &= 1 - \Pr\tup{ X^n \in \bigcup_{s \in \cS} \bigcup_{k \in \set{1,2}} \set{
    \begin{array}{l l}
      x^n \in \cX^n : & I(\type{P}_{x^n}, W_{k,s}) \leq R_k - \epsilon
    \end{array}
  } } \nonumber\\
  &\geq 1 - \sum_{s \in \cS} \sum_{k \in \set{1,2}} \Pr \tup{ X^n \in \set{
    \begin{array}{l l}
      x^n \in \cX^n : & I(\type{P}_{x^n}, W_{k,s}) \leq R_k - \epsilon \\
    \end{array}
  } } \nonumber\\
  &> 1 - \sum_{s \in \cS} \sum_{k \in \set{1,2}} 
        \tup{ \kappa + \exp\tup{e^{-nR/2}} } \nonumber\\
  &= 1 - 2 \abs\cS \tup{ \kappa + \exp\tup{-e^{nR/2}} }.
\end{align}
Since \(\kappa < 1/(2\abs\cS)\), the right-hand side is
positive for sufficiently large \(n\).
We deduce that there exists a sequence \(x^n\in\cX^n\) such that
\(I(\type{P}_{x^n}, W_{k,s}) > R_k - \epsilon\), for every receiver \(k \in \set{1,2}\)
and every state \(s \in \cS\), for \(\lambda < 1/(4\abs\cS)\).
Therefore, let \(\bar{X}\) be distributed according to the type of this sequence,
i.e. \(\bar{X}\sim \hat{P}_{x^n}\), and let \(Y_{1,s},Y_{2,s}\) denote the corresponding outputs.
Then, the rates satisfy
\begin{subequations}
\begin{align}
  R_1 &\leq \min_{s \in \cS} I(\bar{X}; Y_{1,s}), \\
  R_2 &\leq \min_{s \in \cS} I(\bar{X}; Y_{2,s}).
\end{align}
\end{subequations}
Furthermore, if the input constraint is satisfied,
then $\expect [\gamma(\bar{X})] = \frac{1}{n} \sum_{t = 1}^n \gamma(x_t) \leq \Gamma$.
Therefore, we have
\begin{equation}
  \capSAID(\cB)
  \subseteq \scrR(\cB).
\end{equation}
\qed

\section{Proof of Theorem \ref{thm:capSAID-CBC.MIMO}}
\label{sec:proof.capSAID-CBC.MIMO}

We obtain the ID capacity region for MIMO Gaussian channels by approximating
them by discrete channels. This has been done for continuous single-user channels
with certain properties in
\cite{burnashev1999id_approximation_continuous_channel,burnashev2000id_approximation_finite_channel,burnashev2000id_continuous_tit,han2003infoSpectrum_book}
Based on these results, Labidi, Deppe and Boche \cite{labidiDeppeBoche2021ID_mimo}
determined the ID capacity of a single-user MIMO Gaussian channel.
We combine the discretization technique
by Han \cite{han2003infoSpectrum_book} with a continuity argument discussed for
a single-user channel in \cite[Section VI]{peregFerraraBloch2021key_secrecy_bosonicBC_arxiv}
and apply them to the capacity region of the CBC in Theorem \ref{thm:capSAID-CBC}.
Specifically, we consider a discrete channel $\cB^\delta$
that converges to $\cB$ for $\delta \to 0$. We show that
the rate regions $\capSAID(\cB^\delta, \Gamma)$ and $\scrR_\Gamma(\cB^\delta)$
are functions of $\delta$, which are continuous in $\delta = 0$.
To this end, we generalize $\scrR_\Gamma(\cB)$ to the continuous case
by taking the union over all PDFs that satisfy the input
constraint in \eqref{eq:def.R_Gamma}.
We obtain Theorem \ref{thm:capSAID-CBC.MIMO} as the
limit in $\delta \to 0$ of Theorem \ref{thm:capSAID-CBC},
by evaluating this expression for the MIMO Gaussian channel,
following Telatar \cite{telatar1999mimo}.
Thus, we obtain the exact capacity region of the MIMO Gaussian BC.

\subsection{Continuity of \texorpdfstring{$\capSAID(\cB^\delta, \Gamma)$}{C\_ID(B\textasciicircum\delta, \Gamma)}}

Let
$\codet = \set{(\enc[i_1,i_2], \cD_{1,i_1}, \cD_{2,i_2}) : i_1 \in [N_1],\, i_2 \in [N_2]}$
be an $(N_1, N_2, n, \lambda)$ ID-code
for a compound MIMO channel \(\cB = \set{B_s}_{s \in \cS}\).
By \cite[Lemma 6.7.1]{han2003infoSpectrum_book},
there exists a PMF $Q_{i_1,i_2}^\delta$, for every $i_1 \in [N_1]$, $i_2 \in [N_2]$,
and $\delta > 0$, such that the total variation distance
$d(\enc[i_1,i_2] W_{k,s}, Q_{i_1,i_2}^\delta W_{k,s}) \leq \delta$,
for all $k \in \set{1,2}$ and sufficiently large $n$.
Then, since $B_s$ is continuous and smooth,
there exists a CBC \(\cB^\delta = \set{B^\delta_s}_{s \in \cS}\) of
discretized measures \(B^\delta_s\), for every \(s \in \cS\), such that
\begin{equation}
  \lim_{\delta \to 0} B_s^\delta = B_s.
  \label{eq:B_converges_to_Gauss}
\end{equation}
Furthermore, we define discrete decoding sets $\cD_{k,i_k}^\delta$
such that $\lim_{\delta \to 0} \cD_{k,i_k}^\delta = \cD_{k,i_k}$.
The error probabilities are approximated
in terms of the resulting discrete code
$\codet^\delta = \set{(Q^\delta_{i_1,i_2}, \cD_{1,i_1}^\delta, \cD_{2,i_2}^\delta)
  : i_1 \in [N_1],\, i_2 \in [N_2]}$, i.e.
\begin{equation}
  \abs{\saerr_{k,1}(\cB, n, \codet, i_k) -
        \saerr_{k,1}(\cB^\delta, n, \codet^\delta, i_k) }
    < \abs{\saerr_{k,1}(\cB, n, \codet, i_k) -
          \saerr_{k,1}(\cB, n, \codet^\delta, i_k) } + \delta
    < 2\delta,
\end{equation}
and similarly for $\saerr_{k,2}$, for $k \in \set{1,2}$
and sufficiently large $n$.
Hence, by letting $\cB^0 = \cB$ and $\codet^0 = \codet$,
the capacity region $\capSAID(\cB^\delta, \Gamma)$ is continuous in $\delta \geq 0$.

We denote the largest achievable rate \(R_1\) for a given \(R_2\)
and input constraint $\Gamma$ by
\begin{equation*}
  \mathsf{C}_1(\cB^\delta,\Gamma|R_2) = \max\set{R_1 : (R_1, R_2) \in \capSAID(\cB^\delta, \Gamma)}.
\end{equation*}
The equality $\capSAID(\cB^\delta, \Gamma) = \scrR_\Gamma(\cB^\delta)$ implies that
\begin{equation}
  \mathsf{C}_1(\cB^\delta, \Gamma|R_2) = R_1^\star(\cB^\delta, \Gamma|R_2),
  \label{eq:discreteCap_is_R}
\end{equation}
for every \(R_2 \in [0,1]\) and $\delta > 0$.

\subsection{Continuity of \texorpdfstring{$\scrR_\Gamma(\cB^\delta)$}{R\_\Gamma(\cB\textasciicircum\delta)}}

Next, we show that $\scrR_\Gamma(\cB^\delta)$ is also continuous in $\delta = 0$.
Note that \(\scrR_\Gamma(\cB)\) is convex, for every \(\cB\).
To see this, define \(P_{\bar{X}} = a P_X + (1-a) P_{X'}\) for any \(a \in [0,1]\).
If $P_X$ and $P_{X'}$ satisfy
$\max\set{\expect[\gamma(X)], \expect[\gamma(X')]} \leq \Gamma$,
then $\expect[\gamma(\bar{X})] = a \expect[X] + (1-a) \expect[X'] \leq \Gamma$.
Furthermore, by the concavity of $\min$ and the mutual information,
\begin{equation}
  a \min_s I(X; Y_{k,s}) + (1-a) \min_s I(X'; Y_{k,s})
  \leq \min_s I(\bar{X}; Y_{k,s}).
\end{equation}
Therefore, by Slater's condition or, equivalently, by representing
\(\scrR_\Gamma(\cB)\) by its supporting hyperplanes
(see \cite[Theorem 6.20]{rockafellarWets1998variationalAnalysis}),
the optimal $R_1$, given some rate $R_2$, is the optimum
of the Lagrangian function
\begin{equation*}
  R_1^\star(\cB^\delta,\Gamma|R_2) = \min_{\lambda \in \RR} \max_{f_{\bX^\delta} : \expect [\gamma(\bX^\delta)] \leq \Gamma}
     \tup[\Big]{ \min_{s\in\cS} I(\bX^\delta; \bY^\delta_{1,s}) + \lambda \min_{s\in\cS} I(\bX^\delta; \bY^\delta_{2,s}) - \lambda R_2 },
\end{equation*}
and thus $\scrR_\Gamma(\cB^\delta) = \set{ (R_1, R_2) : \displaystyle R_1 \leq R_1^\star(\cB^\delta, \Gamma|R_2) }$.

Since minima and maxima over continuous functions $D \to \RR$
with the same domain $D$ are continuous,
\(R_1^\star(\cB^\delta,\Gamma|R_2)\) is continuous in~$\cB^\delta$.
The composition of two continuous functions is also continuous.
Therefore, \(R_1^\star(\cB^\delta,\Gamma|R_2)\) is a continuous function in~\(\delta\).

\subsection{Equality of \texorpdfstring{$\capSAID(\cB, \Gamma)$ and $\scrR_\Gamma(\cB)$}{C\_ID(B, \Gamma) and R\_\Gamma(B)}}

Theorem \ref{thm:capSAID-CBC} proves for all $\delta > 0$ that
$\capSAID(\cB^\delta, \Gamma) = \scrR_\Gamma(\cB^\delta)$.
Equality holds also in the limit $\delta = 0$, since
\begin{align}
  \mathsf{C}_1(\cB,\Gamma|R_2)
  &\overset{(a)}{=} \mathsf{C}_1(\lim_{\delta \to 0} \cB^\delta , \Gamma|R_2) \nonumber\\
  &\overset{(b)}{=} \lim_{\delta \to 0} \mathsf{C}_1(\cB^\delta , \Gamma|R_2) \nonumber\\
  &\overset{(c)}{=} \lim_{\delta \to 0} R_1^\star(\cB^\delta , \Gamma|R_2) \nonumber\\
  &\overset{(d)}{=} R_1^\star(\lim_{\delta \to 0} \cB^\delta , \Gamma|R_2) \nonumber\\
  &\overset{(a)}{=} R_1^\star(\cB , \Gamma|R_2),
  \label{eq:gaussianCap_is_R}
\end{align}
for every \(R_2 \in [0,1]\), where
(a) follows from the convergence property in \eqref{eq:B_converges_to_Gauss},
(b) holds by the continuity of \(\mathsf{C}_1(\cdot, \Gamma|R_2)\) and the Gaussian
channel measure,
(c) follows from the ID capacity theorem for the compound discrete BC,
Theorem \ref{thm:capSAID-CBC},
and (d) holds by the continuity of \(R_1^\star(\cdot, \Gamma|R_2)\) and the Gaussian
channel measure.
Hence, \eqref{eq:gaussianCap_is_R} implies
\begin{equation}
  \capSAID(\cB,\Gamma) = \scrR_\Gamma(\cB).
\end{equation}

\subsection{Evaluation of the Mutual Information \texorpdfstring{$I(\bX; \bY_{k,s})$}{I(X; Y\_\{k,s\})}}

It remains to show that, for an average power constraint and $\Gamma = P$,
$\scrR_\Gamma(\cB)$ evaluates to the expression \eqref{eq:def.R_P} for $\scrR_P(\cB)$.
The matrix $G_{k,s}^\transp G_{k,s}$ is symmetric and thus has an eigendecomposition
$G_{k,s}^\transp G_{k,s}
  = U_{k,s} \diag(\lambda_{k,s}^{(1)}, \dots, \lambda_{k,s}^{(\tau)}) U_{k,s}^\transp$
with real eigenvalues $\lambda_{k,s}^{(\tau)} \in \RR$ and an orthogonal matrix $U_{k,s}$.
By Telatar \cite{telatar1999mimo}, we have
\begin{equation}
  I(\bX; \bY_{k,s})
    = \frac{1}{2} \log_2 \abs{ G_{k,s} K_\bX G_{k,s}^\transp + I }
    \leq \sum_{j=1}^\tau \frac{1}{2}\log_2(1 + \lambda_{k,s}^{(j)} P_j),
\end{equation}
for \(P_j = (Q_{k,s})_{jj} \geq 0\), \(1 \leq j \leq \tau\), and
\(Q_{k,s} = U_{k,s}^\transp K_\bX U_{k,s}\),
with equality if $Q_{k,s}$ is diagonal.
Furthermore, since $\trace(AB) = \trace(BA)$,
and \(U_{k,s}\) is orthogonal,
\begin{equation}
  \sum_{j=1}^\tau P_j 
    = \trace(U_{k,s}^\transp K_\bX U_{k,s})
    = \trace(K_\bX)
    = \expect[\bX^\transp \bX],
\end{equation}
for \(k \in \set{1,2}\) and \(s \in \cS\).
Hence,
\begin{align}
  \capSAID(\cB, P)
  = \scrR_P(\cB)
    &= 
    \bigcup_{\substack{P_1, \dots, P_\tau: \\ \sum_{j=1}^\tau P_j \leq P}}  \set{
    \begin{array}{l l}
     (R_1, R_2) : & \text{For all $k \in \set{1,2}$: } \\
    		  &\displaystyle R_k \leq \min_{s\in\cS} 
       \sum_{j=1}^\tau \frac{1}{2} \log_2 (1 + \lambda_{k,s}^{(j)} P_j)
    \end{array}
    }.
\end{align}
\qed

\section{Discussion}
\label{sec:discussion}

\subsection{Channel State Information at the Receiver}
\label{sec:capSAID-CBC.CSIR}

In the literature
(e.g.
\cite{weingartenLiuShamaiSteinbergViswanath2007compound_DBC_MIMO_isit,weingartenLiuShamaiSteinbergViswanath2009compound_DBC_MIMO,chongLiang2013compound_DBC,chongLiang2014compound_DBC-K-receiver,chongLiang2014compound_DBC_MIMO_extremal})
it is often assumed that the receivers have perfect knowledge of the channel,
i.e. they know which \(s \in \cS\) was selected.
However, CSI at the receiver does not improve the capacity for compound channels,
since the receiver can estimate the state from a small training sequence \cite[Remark 7.1]{elgamalKim2011network_it}.

\subsection{Pool-Selection vs. Binning}

The pool-selection construction by Bracher and Lapidoth \cite{bracherLapidoth2017idbc_arxiv,bracher2016PhD}
(see Section \ref{sec:proof.capSAID-CBC.achiev})
has similarities with traditional binning, as used in \cite{marton1979DMBC},
and with the binning scheme for identification in \cite{ahlswedeDueck1989id1}.

In traditional binning, Receiver \(k\) assigns randomly a code word \(u^n_k(v_k)\) to every
index from an index set, \(v_k \in \cV_k\), and the index set \(\cV_k\) is
partitioned into disjoint index bins \(\cV_{k,i} \subset \cV_k\) of a specified size.
The set \(\set{u_k^n(v_k) : v_k \in \cV_k}\) must be a reliable code book for
the marginal channel \(W_k\). To simultaneously bin for
two receivers, one transmits a sequence \(x^n(u_1^n, u_2^n)\).

In contrast, in the pool-selection construction, there is only one
set \(\cV\) of indices \(v \in \cV\), and a pool \(\set{x^n(v) : v \in \cV}\).
It may be larger than any reliable code book.
Then, the \(v \in \cV\) are assigned to index bins \(\bcV_{k,i}\) at random, for Receiver \(k\),
and the bin sizes are bounded stochastically (see
Lemma \ref{lemma:BL-binning.cardinalities}) such that every bin
\(\set{x^n(v) : v \in \bcV_{k,i}}\) is effectively a reliable transmission
code book, with high probability.
The sizes of the index sets are chosen such that every bin for Receiver 1
overlaps every bin for Receiver 2 with high probability,
and therefore we can simultaneously
encode for two receivers without needing intermediate code words \(u_1^n, u_2^n\)
that would require auxiliary variables in the random code construction.
However, this construction yields no improvement for transmission,
as a receiver can only reliably distinguish pairwise between bins,
while for transmission one needs to distinguish between
all bins simultaneously, and therefore the union of the bins, the pool, must be
a reliable code book. Then, to transmit over two different channels,
one needs two pools of different sizes, which leads to the traditional binning
scheme.

Next, we compare the pool-selection construction to the ID code construction
in \cite{ahlswedeDueck1989id1}. There, a transmission code book is
divided into overlapping bins, and each bin corresponds to one ID message.
Hence, the pool is a transmission code book. Thus, in a broadcast setting,
the size of the pool is constrained by the capacity of the worse
marginal channel, since we have one pool that must be a reliable transmission
code book for both marginal channels. As the number of possible ID messages
is approximately $2^M$ for bins of size $M$ \cite{ahlswedeDueck1989id1} and
the bins must be smaller than the pool, this construction cannot achieve
the ID capacity for semi-average errors, if the marginal channels admit different
rates for the same input distribution.

\subsection{Identification with Maximal Errors}
\label{sec:discussion.maxID}
It is well-known that in order to identify one establishes common randomness
between a sender and a receiver \cite{ahlswedeDueck1989id2,ahlswedeCsiszar1998CR2}.
In this work, we considered a semi-average error criterion.
As pointed out in Remark \ref{remark:determinsticEncoding}, one
doesn't need local randomness at the sender to establish common randomness
with one receiver, because the messages for the other receiver can be used
as local randomness, i.e. they randomize the channel input.

Under a maximum-error criterion, it is impossible to extract randomness
from the message, since
one has to identify reliably, regardless of the message that was sent to the other receiver.
Hence the error probabilities are maximized over all message \emph{pairs} \((i,j) \in [N_1] \times [N_2]\).
In this setting, Ahlswede \cite{ahlswede2008gtid_updated}
claimed to have proven the exact ID capacity region, yet Bracher and Lapidoth
\cite{bracherLapidoth2017idbc_arxiv,bracher2016PhD}
pointed out a gap in the converse proof.
Nevertheless, Ahlswede \cite{ahlswede2008gtid_updated} showed that
the ID capacity region is the same for two private messages as for
degraded message sets, where one receiver must identify two
messages with rates \(R_0, R_1\), while the other must identify only one of
them with rate \(R_0\). Furthermore, the ID capacity region is
strictly larger than the transmission capacity region with degraded
message sets \cite{ahlswede2008gtid_updated},
because the encoder randomly selects two messages from
sets of cardinalities \(2^{nR_0}\) and \(2^{nR_1}\) and sends both of them over the better channel.
Hence, the receiver can decode a random message from a set
of cardinality \(2^{nR_0} \cdot 2^{nR_1}\), and thus, the ID
rate is \(R_0 + R_1\).

Therefore, using maximal errors, one can achieve a rate region \cite[Theorem 11]{ahlswede2008gtid_updated}
\begin{equation}
  \scrR^{\mathrm{max}}(\cB) = \scrR_1(\cB) \cap \scrR_2(\cB),
\end{equation}
where
\begin{align*}
  \scrR_1(\cB) &= \set{
  \begin{array}{l l}
    (R_1, R_2) : & R_1 \leq \min\set{I(X; Y_1), I(X; Y_1|U) + I(U; Y_2)}, \\
       	         & R_2 \leq I(U; Y_2), \\
		 & \|U\| \leq \abs{\cX} + 2
  \end{array}} \\
  \scrR_2(\cB) &= \set{
  \begin{array}{l l}
    (R_1, R_2) : & R_1 \leq I(U; Y_1), \\
       	         & R_2 \leq \min\set{I(X; Y_2), I(X; Y_2|U) + I(U; Y_1)}, \\
		 & \|U\| \leq \abs{\cX} + 2
  \end{array}}.
\end{align*}
The converse proof is still open, as described above.

\subsection{Open Problems}

We mention a few related open problems.
First, it is of high practical importance to determine
optimal coding strategies for the MIMO BC as
is done for transmission
\cite{palomarCioffiLagunas2003compound_mimo_power,wieselEldarYoninaShamai2007mimoCompoundCapacity_optim,denicCharalambousDjouadi2009compoundMimo_bounds,compoundMIMO_additive_uncertainty,loykaCharalambous2015matrixSVD_ineqs,al-aliHo2017MIMO_precoding_optim,al-asadiEA2019worstCase_beamforming,weingartenSteinbergShamai2006gaussian_mimo_BC,weingartenLiuShamaiSteinbergViswanath2007compound_DBC_MIMO_isit,weingartenLiuShamaiSteinbergViswanath2009compound_DBC_MIMO,benammarPiantanidaShamai2014compound_BC_dirtyPaper,benammarPiantanidaShamai2020compound_BC}.
Furthermore, different models of channel uncertainty can be
considered. The authors are working on a follow-up paper
extending the results of this work to the arbitrarily varying channel,
where the state may change during one transmission/identificaton.
For transmission, such extensions use Ahlswede's Robustification
\cite{ahlswede1980coloringII,ahlswede1986avc_senderKnowsStates,ahlswedeCai1991pinskerAVC}
and Elimination \cite{ahlswede1978avc_elimination} techniques.
However, they cannot be applied directly to ID,
since robustification extends the bound for \emph{one} error probability to
the arbitrarily varying setting, whereas for identification,
one error probability per ID message pair \(i \neq i'\) has to be bounded.
Furthermore, elimination does not apply in the general form of transmission,
where one can prove that for every \emph{random code} there exists a
deterministic code with randomized encoding that achieves the
same rate. This does not apply, since common randomness between
the sender and the receiver, such as a random code, increases the ID
capacity \cite{ahlswedeDueck1989id2,ahlswedeCsiszar1998CR2}.
In practice, one often encounters stochastic channel uncertainty. Appropriate
performance measures should also be considered for ID,
such as ergodic capacities for fast fading and outage capacities for slow fading.
A lower bound on the outage ID capacity has been determined for single-user
channels \cite{ezzineWieseDeppeBoche2021outage_CR}.
For deterministic identification over single-user channels,
without randomized encoding, lower and upper bounds
on the ergodic capacity are known
\cite{salariseddighPeregBocheDeppe2021di_fading,salariseddighPeregBocheDeppe2021det_ID_powerConstraints_tit}.
The latter setting is quite different from the randomized setting considered in this work.
The capacity regions for ID with maximal errors over BCs with fading
can be lower-bounded by the respective transmission capacities,
as was done for single-user channels \cite{ahlswedeCsiszar1998CR2,bocheDeppe2018secureId_wiretap}.
But in contrast to those results, ID converse proofs for BCs seem to be difficult,
especially for maximal errors (see also Section \ref{sec:discussion.maxID}).

\section{Summary}
\label{sec:summary}

We determined the ID capacity region
of the compound BC under a semi-average error criterion
in the MIMO Gaussian setting (Theorem \ref{thm:capSAID-CBC.MIMO}
and in the discrete setting (Theorem \ref{thm:capSAID-CBC}).
To this end, we extended the proofs for the DMBC in
\cite{bracherLapidoth2017idbc_arxiv,bracher2016PhD}
to the compound setting.
In the achievability proof in Section \ref{sec:proof.capSAID-CBC.achiev},
we extended the pool-selection construction
of \cite{bracherLapidoth2017idbc_arxiv,bracher2016PhD}
by modified the decoding sets, specifically
by taking the union over all possible instances of the channel state.
In the converse proof in Section \ref{sec:proof.capSAID-CBC.converse},
we extended the union bound arguments that were
used by Bracher and Lapidoth \cite{bracherLapidoth2017idbc_arxiv,bracher2016PhD}
to combined the rate constraints for the marginal channels.
We applied them to combine bounds for the different states
as well, and thereby showed the existence an input distribution that
satisfies the mutual information constraints on the rates,
for every state.
Our capacity theorem holds for the general compound BC,
as channel ordering conditions, such as degradedness, are not required
in the ID setting.

As examples,
we derived explicit expressions for the ID capacity
regions for symmetric channels in Section \ref{sec:examples.symmetric},
for the binary erasure channel in Section \ref{sec:examples.composedSymmetric},
and for the scalar Gaussian channel in Section \ref{sec:examples.gaussian},
where the sender has no channel state information.
In those examples, each user can achieve the capacity of the
respective marginal channel. Thereby, the capacity region is rectangular
and strictly larger than the transmission capacity region.
However, the ID capacity region is not rectangular in general,
as demonstrated by the example of a compound broadcast Z-channel with a binary state
in Section \ref{sec:examples.z-channel}, and by the example of a
Gaussian product BC in Section \ref{sec:examples.MIMO}.

The shape of the ID capacity regions emphasizes a different behavior
than in the single-user setting (see Section \ref{sec:relatedWork}).
There, the ID capacity equals the
transmission capacity \cite{watanabe2021idMinimaxConverse}.
In the broadcast setting, however, the ID capacity can be strictly larger,
since interference between receivers can be seen as part of
the randomization of the coding scheme, as discussed in 
Remark \ref{remark:determinsticEncoding} and Section \ref{sec:discussion.maxID}.

\section{Acknowledgments}

The authors would like to thank Holger Boche for pointing out a gap in the original
continuity argument in the proof of Theorem \ref{thm:capSAID-CBC.MIMO}.
This work was supported by the German Ministry of Education and Research (BMBF) 
through grants
16KISK002 (Johannes Rosenberger, Christian Deppe), 
16KISQ028 (Uzi Pereg, Christian Deppe), 
16KIS1005 (Christian Deppe).

\ifarxivstrict
\newcommand{\arxivHref}[1]{\href{https://arxiv.org/abs/#1}{#1}}

\else
  \bibliography{IEEEabrv,short,bib}
\fi

\end{document}